\documentclass[aps,graphicx,twocolumn]{revtex4}

\usepackage{graphicx}
\usepackage{epstopdf}
\usepackage{verbatim}
\usepackage{amsmath}
\usepackage[colorlinks,linkcolor=blue,citecolor=blue,urlcolor=blue]{hyperref}
\usepackage[caption=false,font=scriptsize,labelfont=sf,textfont=sf]{subfig}
\usepackage{multirow}

\begin{document}
\title{Device-independent quantum secret sharing with advanced random key generation basis}

\author{Qi Zhang$^{1}$, Jia-Wei Ying$^{2}$, Zhong-Jian Wang$^{1}$, Wei Zhong$^{3}$, Ming-Ming Du$^{2}$, Shu-Ting Shen$^{2}$,\\
Xi-Yun Li$^{1}$, An-Lei Zhang$^{1}$, Shi-Pu Gu$^{2}$, Xing-Fu Wang$^{1}$, Lan Zhou$^{1}$\footnote{Email address: zhoul@njupt.edu.cn}, Yu-Bo Sheng$^{2}$}
\address{
$^1$College of Science, Nanjing University of Posts and Telecommunications, Nanjing, Jiangsu 210023, China\\
$^2$College of Electronic and Optical Engineering and College of Flexible Electronics (Future Technology), Nanjing University of Posts and Telecommunications, Nanjing, Jiangsu 210023, China\\
$^3$Institute of Quantum Information and Technology, Nanjing University of Posts and Telecommunications, Nanjing, Jiangsu 210003, China\\
}
\date{\today}

\begin{abstract}
Quantum secret sharing (QSS) enables a dealer to securely distribute keys to multiple players. Device-independent (DI) QSS can resist all possible attacks from practical imperfect devices and provide QSS the highest level of security in theory. However, DI QSS requires high-performance devices, especially for low-noise channels, which is a big challenge for its experimental demonstration. We propose a DI QSS protocol with the advanced random key generation basis strategy, which combines the random key generation basis with the noise preprocessing and postselection strategies. We develop the methods to simplify Eve's conditional entropy bound and numerically simulate the key generation rate in an acceptable time. Our DI QSS protocol has some advantages. First, it can increase the noise tolerance threshold from initial 7.147\% to 9.231\% (29.16\% growth), and reduce the global detection efficiency threshold from 96.32\% to 93.41\%. The maximal distance between any two users increases to 1.43 km, which is about 5.5 times of the initial value. Second, by randomly selecting two basis combinations to generate the key, our DI QSS protocol can reduce the entanglement resource consumption. Our protocol has potential for DI QSS's experimental demonstration and application in the future.
\end{abstract}
\maketitle

\section{Introduction}\label{Section1}
Quantum communication aims to guarantee the unconditional communication security based on the fundamental principles of quantum mechanics. As a multipartite cryptographic primitive \cite{QSS1,QSS2,QSS3}, quantum secret sharing (QSS) is one of the main quantum communication branches. QSS divides the key of the dealer into multiple parts and distributes each part to a player. Any subset of players cannot reconstruct the distributed key, which can only be reconstructed when all the players cooperate \cite{QSS1}.

The first QSS protocol based on the Greenberger-Horne-Zeilinger (GHZ) state was proposed in 1999 \cite{QSS1}. Since then, QSS has been widely studied both in theory and experiment \cite{QSS2,QSS3,QSS4,QSS5,QSS6,QSS7,QSS8,QSS9,QSS10,MDIQSS1,MDIQSS2,MDIQSS3,QSS11,QSS12,QSS13,QSS14,QSS15,QSS16,QSS17,QSS18,QSS19,QSS20,QSS21}. Like other quantum communication branches, such as quantum key distribution (QKD), the practical imperfect experimental devices may cause serious security loopholes \cite{cryptography1,cryptography2,cryptography3}. The device-independent (DI) paradigm provides a promising method to guarantee the security based on the nonlocal correlation of the measurement results \cite{nonlocal1}. DI type protocols can defend against all possible attacks from practical experimental devices and provide the highest level of security \cite{DIrandom1,DIrandom2,DIQKD1,DIQKD2,DIQKD3}. The research of DI-type protocols started in the DI QKD system \cite{DIQKD1,DIQKD2,DIQKD3}. Since 2020, DI theory have been introduced into the quantum secure direct communication field \cite{DIQSDC1,DIQSDC2,DIQSDC3}. In 2022, DI QKD's experiments have achieved significant breakthroughs \cite{DIQKDe1,DIQKDe2,DIQKDe3}, thus proving the feasibility of the DI theory. The research on DI QSS has been started since 2019. Refs.~\cite{DIQSS2,DIQSS3} proved DI QSS's correctness and completeness under a causal independence assumption regarding measurement devices. In 2024, the DI QSS protocol based on the violation of Svetlichny inequality was proposed \cite{DIQSS1}. This protocol first characterized DI QSS's performance in a practical noisy communication scenario by numerically simulating its key generation rate, global detection efficiency threshold, noise threshold, and the maximal communication distance between any two users. Similar as DI QKD protocols \cite{DIQKD2,DIQKD3}, DI QSS protocol requires quite high global detection efficiency and has low noise resistance \cite{DIQSS1}. In particular, as DI QSS requires to distribute multi-photon entanglement to multiple users while DI QKD only requires to distribute EPR state to two users, DI QSS is more vulnerable to channel loss and limited detection efficiency. It requires ultra-low-noise settings with the global detection efficiency as high as 96.32\% and the noise threshold of 7.148\% \cite{DIQSS1}.

In the DI QKD field, some active improvement strategies have been put forward to enhance its noise resistance and reduce the high detection efficiency requirement, including noise preprocessing \cite{DIQKD9,DIQKD10,DIQKD11}, postselection \cite{DIQKD10,DIQKD11,DIQKD12,DIQKD13}, random key generation basis \cite{DIQKD12,DIQKD15}, and advantage distillation \cite{DIQKD17}. For example, the adoption of the noise preprocessing strategy increase DI QKD's noise threshold from initial 7.1492\% to 8.0848\% \cite{DIQKD10,DIQKD12}. By combining the random postselection and noise preprocessing strategies, DI QKD's global detection efficiency threshold was further reduced to less than 87.49\%. Benefit for the improvement strategies, the first DI QKD experiment in the optical platform is experimentally demonstrated \cite{DIQKDe3}. In 2021, the DI QKD protocol with random key generation basis strategy was proposed \cite{DIQKD15}, which effectively improved the noise tolerance threshold from the initial 7.15\% to 8.24\%. However, the estimation of Eve's conditional entropy bound is quite complicated, which requires about 5000 core hours and thus largely increases the difficulty of parameter optimization. Soon later, Ref.~\cite{DIQKD12} simplified the DI conditional entropy bound and provided the optimal entropy in a much shorter time (less than one second). It largely reduced the difficulty of parameter optimization and further improved DI QKD's noise tolerance threshold to 8.36\%. The noise preprocessing combined with the random random postselection strategy was also introduced into the DI QSS protocol \cite{DIQSS1}, which can reduce its global detection efficiency threshold from 96.32\% to 94.30\% and increase the noise threshold from 7.148\% to 8.072\%.

Although the above improvement can reduce DI QSS's experimental difficulty, the experimental demonstration of DI QSS is still a big challenge. Meanwhile, the DI QSS protocol \cite{DIQSS1} only uses one basis combination to generate the key and discards the measurement results for the other basis combination, which is a waste of precious entanglement resources. For further enhancing DI QSS's noise resistance, and reducing the global detection efficiency and the entanglement resource consumption, we propose the DI QSS protocol with advanced random key generation basis strategies, which combine the random key generation basis, noise preprocessing, and postselection strategies. We develop the technique in Ref.~\cite{DIQKD12} to simplify Eve's conditional entropy bound and numerically simulate its performance in the practical noise environment. The advanced random key generation basis strategy can increase DI QSS's noise tolerance threshold from 7.147\% to 9.231\% (29.16\% growth) and reduce the global detection efficiency requirement from initial 96.32\% to 93.41\% ($q\rightarrow0.5$). In this way, the maximal secure communication distance between any two users can be increased from 0.26 km to 1.43 km (about 5.5 times). Based on the above features, our DI QSS protocol with advanced random key generation basis strategy can effectively facilitate DI QSS's experimental implementations and application in the future quantum communication field.

The paper is organized as follows. In Sec.~\ref{Section2}, we briefly describe the tripartite joint key generation process and the detection of genuine tripartite nonlocal correlations. In Sec.~\ref{Section3}, we present the DI QSS protocol with the random key generation basis. In Sec.~\ref{Section4}, we reduce the conditional entropy bound problem of the two-basis variant to qubit system to estimate its key generation rate in the practical noisy channel environment. In Sec.~\ref{Section5}, we combine the noise preprocessing strategy to further enhance DI QSS's resistance to channel noise. Finally, we make some discussion and draw a conclusion in Sec.~\ref{Section6}.

\section{Tripartite joint key generation and genuine tripartite nonlocal correlation detection}\label{Section2}
In the DI QSS, three users' devices are successively repeated for $n$ ($n\rightarrow\infty$) rounds. Their measurement results have two purposes: (1) some rounds of measurement results are used to generate the joint key bits and estimate the qubit error rate (QBER); (2) some rounds of measurement results are used to estimate the genuine tripartite nonlocal correlation among three users' results for security checking.

\subsection{Tripartite joint key generation with the random key generation basis strategy}\label{Section2.1}
In this subsection, we explain the joint key generation rules of our DI QSS protocol with the random key generation basis strategy. Suppose that three users Alice, Bob and Charlie share a three-photon GHZ state with the form of
\begin{eqnarray}\label{GHZ1}
|GHZ\rangle=\frac{1}{\sqrt{2}}\left(|HHH\rangle+|VVV\rangle\right),
\end{eqnarray}
where $|H\rangle$ and $|V\rangle$ denote the horizontal and vertical polarization of the photon, respectively.

Three users independently and randomly perform measurement on each photon in their own locations with the bases $A_i$, $B_j$, and $C_k$, respectively, where $i,k\in\{1,2\}$ and $j\in\{1,2,3\}$. In detail, Alice has two measurement bases $A_1=\sigma_x$, $A_2=\sigma_y$, Bob has three measurement bases $B_1=\sigma_x$, $B_2=\frac{\sigma_x-\sigma_y}{\sqrt{2}}$, $B_3=\frac{\sigma_x+\sigma_y}{\sqrt{2}}$, and Charlie has two measurement bases $C_1=\sigma_x$, $C_2=-\sigma_y$, where $\sigma_x$ and $\sigma_y$ are Pauli operators with the eigenstates as
\begin{eqnarray}\label{eigenstate1}
|+_x\rangle&=&\frac{1}{\sqrt{2}}\left(|H\rangle+|V\rangle\right),
|-_x\rangle=\frac{1}{\sqrt{2}}\left(|H\rangle-|V\rangle\right),\nonumber\\
|+_y\rangle&=&\frac{1}{\sqrt{2}}\left(|H\rangle+i|V\rangle\right),
|-_y\rangle=\frac{1}{\sqrt{2}}\left(|H\rangle-i|V\rangle\right).\nonumber\\
\end{eqnarray}
Their binary measurement results are denoted as $a_i,b_j,c_k\in\{-1,+1\}$, respectively, where $|+_x\rangle$ and $|+_y\rangle$ are labeled as $+1$, while $|-_x\rangle$ and $|-_y\rangle$ are labeled as $-1$.

Two measurement basis combinations $\{A_{1}B_{1}C_{1}\}$ and $\{A_{2}B_{1}C_{2}\}$ are used for the joint key generation. The key generation includes two cases.

Case 1: The users measure the three photons in a GHZ state with the basis combination $\{A_{1}B_{1}C_{1}\}$. They can obtain
\begin{eqnarray}\label{case1}
|GHZ\rangle&=&\frac{1}{2}\Big(|+_x\rangle_a|+_x\rangle_b|+_x\rangle_c+|+_x\rangle_a|-_x\rangle_b|-_x\rangle_c \nonumber\\
&+&|-_x\rangle_a|+_x\rangle_b|-_x\rangle_c+|-_x\rangle_a|-_x\rangle_b|+_x\rangle_c\Big).
\end{eqnarray}

Case 2: The users measure three photons with the basis combination $\{A_{2}B_{1}C_{2}\}$, where
\begin{eqnarray}\label{case2}
|GHZ\rangle&=&\frac{1}{2}\Big(|+_y\rangle_a|+_x\rangle_b|+_y\rangle_c+|+_y\rangle_a|-_x\rangle_b|-_y\rangle_c \nonumber\\
&+&|-_y\rangle_a|+_x\rangle_b|-_y\rangle_c+|-_y\rangle_a|-_x\rangle_b|+_y\rangle_c\Big).
\end{eqnarray}

In above cases, the relationship among the users' measurement results is summarized in Tab.~\ref{table1}. If Charlie knows Alice's and Bob's measurement bases, he can know whether Alice's and Bob's measurement results are identical or opposite, but he cannot explicitly know Alice's and Bob's specific measurement results.

\begin{table}
\includegraphics[scale=0.43]{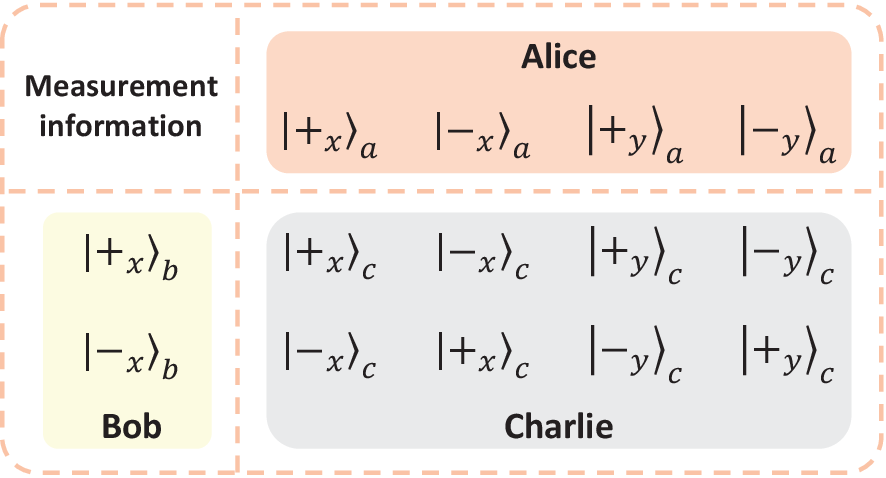}
\caption{(Color online) Relationships of three users' measurement results. Alice's measurement results are shown in the first row, and Bob's measurement results are shown in the first column. Charlie's measurement results appear in the boxes.}
\label{table1}
\end{table}

We label the measurement result $+1$ as the key bit 0, and $-1$ as 1. The relationship among the measurement results in Tab.~\ref{table1} can be summarized as $K_A=K_B\oplus K_C$, where $K_A$, $K_B$, and $K_C$ denote the key bits of Alice, Bob, and Charlie, respectively.

\subsection{Genuine tripartite nonlocal correlation detection}\label{Section2.2}
The genuine tripartite nonlocality can be detected by the Bell-type inequality for three-qubit systems, called the Svetlichny inequality \cite{Svetlichny}. With the above measurement bases in Sec. ~\ref{Section2.1}, the Svetlichny polynomial $S_{ABC}$ can be written as
\begin{eqnarray}\label{Svetlichnypolynomial}
S_{ABC}&=\langle a_{1}b_{2}c_{2}\rangle+\langle a_{1}b_{3}c_{1}\rangle+\langle a_{2}b_{2}c_{1}\rangle-\langle a_{2}b_{3}c_{2}\rangle \nonumber\\
&+\langle a_{2}b_{3}c_{1}\rangle+\langle a_{2}b_{2}c_{2}\rangle+\langle a_{1}b_{3}c_{2}\rangle-\langle a_{1}b_{2}c_{1}\rangle.
\end{eqnarray}
Here, $\langle a_{i}b_{j}c_{k}\rangle$ is the expected value of the tripartite measurement results with the form of  $\langle a_{i}b_{j}c_{k}\rangle =P(a_{i}b_{j}c_{k}=1)-P(a_{i}b_{j}c_{k}=-1)$ ($P$ represents the probability). If the three photons are in classical correlation,  $S_{ABC}$ satisfies the Svetlichny inequality $S_{ABC}\leq4$. The violation of Svetlichny inequality ($S_{ABC}>4$) implies the existence of the genuine tripartite nonlocality.

For the three-photon GHZ state,  $S_{ABC}$ can be simplified by the CHSH polynomials as \cite{nonlocal4}
\begin{eqnarray}\label{Svetlichnypolynomial2}
S_{ABC}=\langle S_{AB}c_{2}\rangle+\langle S_{AB}'c_{1}\rangle.
\end{eqnarray}
Here, $S_{AB}$ and $S_{AB}'$ are the general CHSH polynomial and its equivalent variation between Alice's and Bob's measurement results with the form of
\begin{eqnarray}\label{CHSH}
S_{AB}&=&\langle a_{1}b_{2}\rangle+\langle a_{2}b_{2}\rangle+\langle a_{1}b_{3}\rangle-\langle a_{2}b_{3}\rangle,\nonumber\\
S_{AB}'&=& \langle a_{2}b_{3}\rangle+\langle a_{2}b_{2}\rangle+\langle a_{1}b_{3}\rangle-\langle a_{1}b_{2}\rangle,
\end{eqnarray}
with $\langle a_{i}b_{j}\rangle=P(a_{i}b_{j}=1)-P(a_{i}b_{j}=-1)$.

For simplicity, we define $S$ to represent $S_{AB}$ or $S_{AB}'$ based on the following rule
 \begin{equation}\label{Bell}
	S=\begin{cases}
	   S_{AB}' ,& \text{if     $c_1=+1$},\\
	   -S_{AB}' ,& \text{if     $c_1=-1$},\\
        S_{AB} ,& \text{if     $c_2=+1$},\\
	   -S_{AB} ,& \text{if     $c_2=-1$}.
	\end{cases}
\end{equation}
With the definition, the violation of Svetlichny inequality equals to Alice's and Bob's measurement results violating the CHSH inequality ($S>2$) \cite{nonlocal4}.

\section{DI QSS protocol with random key generation basis strategy}\label{Section3}
Here, we explain our DI QSS protocol with random key generation basis strategy. Similar to previous DI protocols \cite{DIQKD1,DIQKD2,DIQKD3,DIQSDC1,DIQSDC2,DIQSDC3,DIQKD4,DIQKD5,DIQKD6}, the security of this DI QSS protocol can be guaranteed by two basic assumptions, say, the quantum physics is correct and all users' physical locations are secure. Meanwhile, the three users must be legitimate and honest. Three users' devices are repeatedly used for $n$ ($n\to \infty $) rounds. The schematic diagram of DI QSS protocol with random key generation basis strategy is shown in Fig.~\ref{fig1}, which includes the following five steps:

\begin{figure}
\includegraphics[scale=0.125]{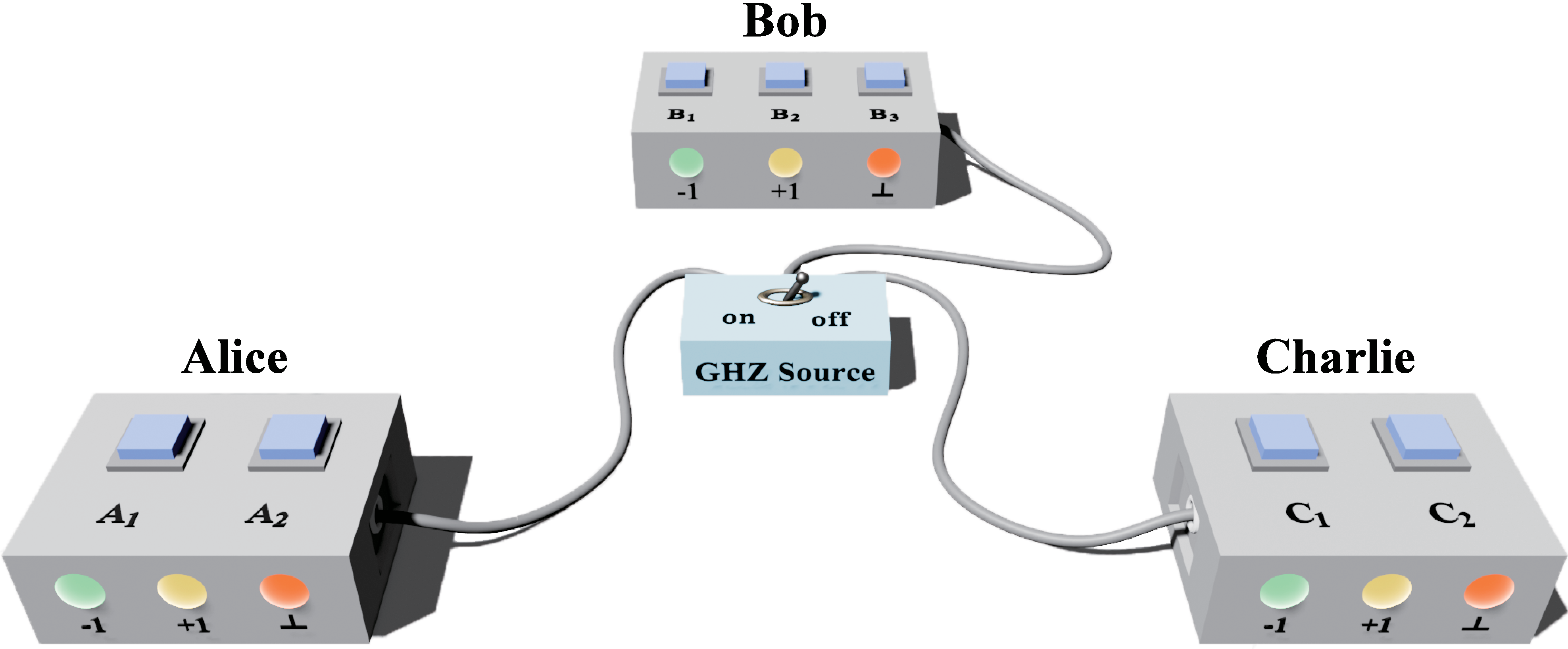}
\caption{(Color online) Schematic diagram of the DI QSS protocol with random key generation basis strategy. The central entanglement source prepares $n$ identical three-photon GHZ states, which are split into three sequences, namely $E_1$, $E_2$, and $E_3$ sequences. The photons in three sequences are distributed to three remote users, Alice, Bob, and Charlie, respectively. The three users independently select the measurement settings ($i$, $j$ and $k$) and record the measurement results: $-1$, $+1$, or no-click event $\perp$.}
\label{fig1}
\end{figure}

\textbf{Step 1 Entanglement distribution.} The central entanglement source prepares $n$ pairs of three-photon polarization GHZ states \eqref{GHZ1}. The source divides the three photons in each GHZ state into $E_1$, $E_2$, and $E_3$ sequences. All the photons in the three sequences are sequentially transmitted to Alice, Bob, and Charlie, respectively.

\textbf{Step 2 Measurement basis selection.} Three users independently and randomly select the measurement bases to measure all photons in the sequences $E_1$, $E_2$, and $E_3$, respectively. The measurement bases $A_i$, $B_j$, and $C_k$ are described in Sec.~\ref{Section2.1}, where $i,k\in\{1,2\}$ and $j\in\{1,2,3\}$. The tripartite measurement results are denoted as $a_i$, $b_j$, and $c_k$, where $a_i,b_j,c_k\in\{-1,+1\}$. The detector's no-click event $\perp$ is randomly noted as $-1$ or $+1$, which is called the postselection operation.

\textbf{Step 3 Parameter estimation.} After all measurements, Alice, Bob, and Charlie sequentially announce the measurement basis for each photon. According to the measurement basis selection, the parameter estimation can be categorized into three cases.

Case 1: Bob chooses the measurement basis $B_2$ or $B_3$. In this case, all three users announce the measurement results, which are used to estimate the Svetlichny polynomial $S_{ABC}$ and the CHSH polynomial $S$. When $2<S\leq 2\sqrt{2}$ (equivalent to $4<S_{ABC}\leq 4\sqrt{2}$), the corresponding results at three users' locations have genuine nonlocal correlation. The user can bound the key leakage rate to Eve, thus the photon transmission process is secure and the communication continues. When $S\leq 2$ (equivalent to $S_{ABC}\leq 4$), the measurement results are only classical correlated. The users cannot detect Eve's attacking. In this way, the users regard that the photon transmission process is not secure, and the communication should be terminated.

Case 2: When the measurement basis combination is $\{A_{1}B_{1}C_{1}\}$ or $\{A_{2}B_{1}C_{2}\}$. Alice randomly announces a part of her measurement results, and Bob (Charlie) announces his corresponding measurement results to estimate the QBER $\delta$.  The users preserve the remaining unannounced measurement results as the raw key bits.

Case 3: When the measurement basis combination is $\{A_{1}B_{1}C_{2}\}$ or $\{A_{2}B_{1}C_{1}\}$, the users have to discard their measurement results.

\textbf{Step 4 Error correction and private amplification.} The users repeat the above process until all the users have obtained a sufficient number of secure key bits. Then, the users perform the error correction and private amplification on the raw key bits, eventually forming a series of secure key bits.

\textbf{Step 5 Secret reconstruction.} Charlie announces his key bit $K_C$, and Bob combines his key bit $K_B$ to reconstruct the key bit $K_A$ delivered by Alice based on the encoding rule in Sec.~\ref{Section2.1}.

\section{The performance evaluation of the DI QSS protocol with random key generation basis strategy}\label{Section4}
In this section, we evaluate the performance of the DI QSS protocol with random key generation basis strategy in the practical noisy environment against collective attacks. We assume that Alice (Charlie) uses measurement basis $A_1$ ($C_1$) and $A_2$ ($C_2$) with probabilities $p$ and $\bar{p}=1-p$, respectively.

The users generate the raw key bits when the measurement basis combination is $\{A_{1}B_{1}C_{1}\}$ or $\{A_{2}B_{1}C_{2}\}$. In the asymptotic limit of a large number of rounds, we calculate the asymptotic key generation rate $r_\infty$, which is the ratio of the extractable key length to measurement rounds $n$ ($n\to\infty$). Taking into account the ``sifting'' effect, the asymptotic key generation rate can be written as \cite{DIQKD12,DIQKD15}
\begin{eqnarray}\label{rinfty1}
r_\infty = p^2 r_{111}+\bar{p}^2 r_{212}= \left(p^2+\bar{p}^2\right)\left(\lambda r_{111}+\bar{\lambda}r_{212}\right),
\end{eqnarray}
where $\lambda=p^2/(p^2+\bar{p}^2)$ represents the matching weight for two key generation basis combinations, and $\bar{\lambda}=1-\lambda$. We adopt the Devetak-Winter bound \cite{DWrate,DWrate2} to estimate the key generation rates corresponding to the measurement basis combination $\{A_1B_1C_1\}$ and $\{A_2B_1C_2\}$. The Devetak-Winter bound is a universal method for calculating the key rate in the quantum cryptography field, which has been widely used in QKD and QSS systems \cite{DWQKD1,DWQKD2,DWQKD3,DWQKD4}, and extended to DI QKD and one-side DI QSS systems under the collective attack assumption \cite{DIQKD9,DIQKD17,DWQSS1,DWQKD5,DWQKD6}. In the asymptotic limit of a large number of rounds, the key generation rates corresponding to the measurement basis combination $\{A_{1}B_{1}C_{1}\}$ and $\{A_{2}B_{1}C_{2}\}$ can be provided by \cite{DWrate,DWrate2,DIQSS1}
\begin{eqnarray}\label{r111212}
r_{111}&=& H\left(A_1|E\right)-H\left(A_1|B_1,C_1\right),\nonumber\\
r_{212}&=& H\left(A_2|E\right)-H\left(A_2|B_1,C_2\right),
\end{eqnarray}
where $H(|)$ is the conditional von Neumann entropy. $H\left(A_1|E\right)$ and $H\left(A_2|E\right)$ estimate the uncertainty of Eve's knowledge about Alice's key bits when he knows Charlie's key bits, say, the key secrecy rate to Eve corresponding to the measurement basis combinations $\{A_{1}B_{1}C_{1}\}$ and $\{A_{2}B_{1}C_{2}\}$, respectively. $H\left(A_1|B_1,C_1\right)$ and $H\left(A_2|B_1,C_2\right)$ quantify the tripartite raw key irrelevance with the measurement basis combinations $\{A_{1}B_{1}C_{1}\}$ and $\{A_{2}B_{1}C_{2}\}$, respectively. Therefore, Eq.~\eqref{rinfty1} is rewritten as
\begin{eqnarray}\label{rinfty2}
r_\infty &=& \left(p^2+\bar{p}^2\right)\Big[\left(\lambda H\left(A_1|E\right)+\bar{\lambda}H\left(A_2|E\right)\right) \nonumber\\
&-&\left(\lambda H\left(A_1|B_1,C_1\right)+\bar{\lambda}H\left(A_2|B_1,C_2\right)\right)\Big].
\end{eqnarray}

\subsection{The total key secrecy rate to Eve}\label{Section4.1}
We apply the similar parameter estimation technique in \cite{DIQKD12} in our DI QSS protocol with random key generation basis strategy. Firstly, we consider the total key secrecy rate to Eve, which is the first term in Eq.~\eqref{rinfty2}. We define the total key secrecy rate to Eve as $H\left(A|E\right)$ with the form of
\begin{eqnarray}\label{keysecrecyrate1}
H\left(A|E\right)=\lambda H\left(A_1|E\right)+\bar{\lambda}H\left(A_2|E\right).
\end{eqnarray}

Under the collective attack model \cite{DIQKD2,DIQKD3,DIQSS1}, the key secrecy rate to Eve is a function of CHSH polynomial $S$ between Alice and Bob. A rough lower bounds on $H\left(A_1|E\right)$ and $H\left(A_2|E\right)$ are given by
\begin{eqnarray}\label{keysecrecyrate2}
H\left(A_1|E\right)=H\left(A_2|E\right)\geq 1-\phi\left(\sqrt{\frac{S^2}{4}-1}\right),
\end{eqnarray}
where $\phi(x)=h(\frac{1}{2}+\frac{1}{2}x)$ and the binary Shannon entropy $h(x)=-x\log_{2}{x}-(1-x)\log_{2}{(1-x)}$. We define the lower bound function $g(x)=1-\phi(x)$, which is a convex function.

In the DI scenario, we consider that all devices are untrusted. In this way, the measurement basis used by three users is potentially under Eve's control, so all measurement bases need to be reparametrized in the following analysis, as described in Appendix \ref{Appendix A}. We use the qubit uncertainty relation to maximally bound the conditional entropy of Alice's key generation. In this way, $H\left(A|E\right)$ is bounded by
\begin{eqnarray}\label{keysecrecyrate3}
H\left(A|E\right)&=&\lambda H\left(A_1|E\right)+\bar{\lambda}H\left(A_2|E\right)\nonumber\\
&\geq & g\left(\sqrt{\lambda\left\langle\bar{A_1}\otimes B\right\rangle^2+\bar{\lambda}\left\langle\bar{A_2}\otimes B'\right\rangle^2}\right).\nonumber\\
\end{eqnarray}
Here, the Pauli observations $\bar{A_1}$ and $\bar{A_2}$ are orthogonal to $A_1$ and $A_2$ in the X-Y plane of the Bloch sphere, respectively. $B$ and $B'$ are observations with any given eigenvalue $\pm 1$ on Bob's subsystem. In DI-type protocols, we cannot directly obtain the correlation between Alice's and Bob's subsystems $\left\langle\bar{A_i}\otimes B\right\rangle$. Inspired by Ref.~\cite{DIQKD12}, Eve's information about the measurement results corresponding to $A_1$ and $A_2$ bases can be maximized by the correlation operator of Bob's subsystem with the virtual complementary $\bar{A_1}$ and $\bar{A_2}$.

Based on the correlations among Alice, Bob, and Charlie, \emph{i.e.}, the CHSH polynomial $S$ (Svetlichny polynomial $S_{ABC}$), we establish a DI qubit correlation bound on $\left\langle\bar{A_i}\otimes B\right\rangle$. This process is explained in detail in Appendix \ref{Appendix A}, which requires to solve the maximization problem as
\begin{eqnarray}\label{maxproblem}
E_\lambda(S)^2=\max& s^2g^2+c^2h^2+2(2\lambda-1)scgh\varDelta, \nonumber\\
s.t.& cg+sh\geq S/2, \nonumber\\
& g^2 \leq 1, \nonumber\\
& h^2 \leq 1, \nonumber\\
& (1-g^2)(1-h^2)\geq g^2h^2\varDelta^2, \nonumber\\
& c^2+s^2=1, \nonumber\\
& \varDelta ^{2} \leq 1,
\end{eqnarray}
where we parameterize the Pauli operator correlation $\langle\sigma_x\otimes\sigma_x\rangle=g\cos\gamma$, $\langle\sigma_x\otimes\sigma_y\rangle=g\sin\gamma$, $\langle\sigma_y\otimes\sigma_x\rangle=h\cos\mu$, $\langle\sigma_y\otimes\sigma_y\rangle=h\sin\mu$, $\varDelta=\cos(\gamma-\mu)$. We obtain the virtual orthogonal operator $\bar{A_i}=(-1)^{i-1}\sin\frac{\varphi_A}{2}\sigma_x+\cos\frac{\varphi_A}{2}\sigma_y$ and $s=\sin\frac{\varphi_A}{2}$, $c=\cos\frac{\varphi_A}{2}$.

Then, the conditional entropy bound of Eq.~\eqref{keysecrecyrate3} is given by
\begin{eqnarray}\label{keysecrecyrate4}
H\left(A|E\right)\ge g\left(\tilde{E}_\lambda(S)\right),
\end{eqnarray}
where $\tilde{E}_\lambda(S)\equiv \sqrt{\tilde{E}_\lambda(S)^2}$, and $\tilde{E}_\lambda(S)^2$ is the maximum value that can be obtained by numerically and analytically solving the optimization problem \eqref{maxproblem}. Finally, a convexity analysis of the DI entropy bound is performed \cite{DIQKD12}.

\begin{figure}
\includegraphics[scale=0.37]{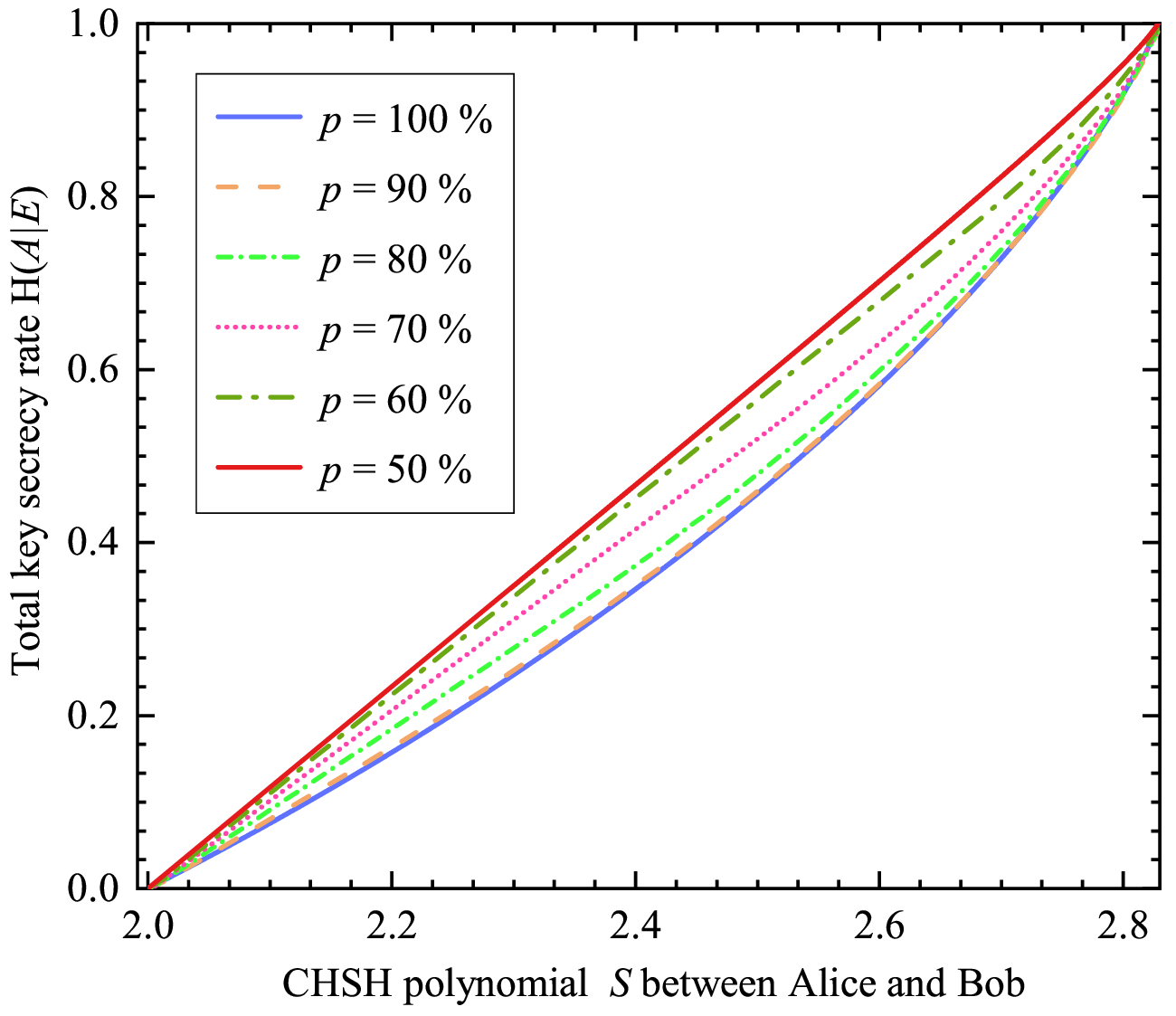}
\caption{(Color online) The total key secrecy rate to Eve $H\left(A|E\right)$ as a function of the CHSH polynomial $S$ between Alice's and Bob's photons. Here, we fix the probability that Alice (Charlie) chooses measure $A_1$ ($C_1$) as $p=100\%$, 90\%, 80\%, 70\%, 60\%, 50\%, respectively. The calculation results corresponding to $p=0$, 10\%, 20\%, 30\%, 40\% are consistent with those corresponding to $p=100\%$, 90\%, 80\%, 70\%, 60\%, respectively.}
\label{fig2}
\end{figure}

In Fig.~\ref{fig2}, we use CHSH polynomial $S$ between Alice's and Bob's measurement results to quantify $H\left(A|E\right)$. For example, with $S=2.4$, when Alice (Charlie) equally ($p=50\%$) uses $A_1$ or $A_2$ ($C_1$ or $C_2$) to measure the photon, $H\left(A|E\right)=0.467$ can be obtained. It is higher than the value of 0.346 corresponding to Alice (Charlie) only using $A_1$ ($C_1$) to generate the key ($p=100\%$). Higher $H\left(A|E\right)$ enables our DI QSS protocol with random key generation basis strategy to have stronger noise resistance.

\subsection{The key generation rate in practical noise model}\label{Section4.2}
During practical long-distance photon transmission, photon loss and decoherence caused by the channel noise are unavoidable, which can severely degrade the entanglement and weaken the nonlocal correlation among users' measurement results. Here, we consider the common white noise model, which has been widely adopted in DI QKD protocols \cite{DIQKD3,DIQKD9,DIQKD10,DIQKD11,DIQKD12,DIQKD13}. In the white noise model, the target GHZ state may degenerate to eight possible GHZ states with equal probability. Meanwhile, we assume that each user successfully detects the transmitted photon with the probability of $\eta$, and the no-click event occurs with the probability of $\bar{\eta}=1-\eta$. In this way, after distributing the entangled photons, three users share $n$ pairs of identical mixed states with the form of
\begin{eqnarray}\label{rhoABC1}
\rho_{ABC}'&=&\eta^3\rho_{ABC}+(1-\eta)^3|vac\rangle\langle vac|\nonumber\\
&+&\frac{1}{2}\eta^2(1-\eta)\left(|HH\rangle\langle HH|+|VV\rangle\langle VV|\right)_{BC}\nonumber\\
&+&\frac{1}{2}\eta^2(1-\eta)\left(|HH\rangle\langle HH|+|VV\rangle\langle VV|\right)_{AC}\nonumber\\
&+&\frac{1}{2}\eta^2(1-\eta)\left(|HH\rangle\langle HH|+|VV\rangle\langle VV|\right)_{AB}\nonumber\\
&+&\frac{1}{2}\eta(1-\eta)^2\left(|H\rangle\langle H|+|V\rangle\langle V|\right)_{A}\nonumber\\
&+&\frac{1}{2}\eta(1-\eta)^2\left(|H\rangle\langle H|+|V\rangle\langle V|\right)_{B}\nonumber\\
&+&\frac{1}{2}\eta(1-\eta)^2\left(|H\rangle\langle H|+|V\rangle\langle V|\right)_{C},
\end{eqnarray}
where $|vac\rangle$ denotes the vacuum state, and
\begin{eqnarray}\label{rhoABC}
\rho_{ABC}=F|GHZ\rangle\langle GHZ|+\frac{1-F}{8} I.
\end{eqnarray}
The fidelity $F$ is the probability that no error occurs in the photonic state, and the unit matrix $I$ is composed by the density matrices of the eight possible noisy GHZ states.

We first consider the case that all photons in the GHZ states are successfully detected by three users. According to the coding rule $K_A=K_B\oplus K_C$, four GHZ states will make three users obtain incorrect measurement results corresponding to the key generation basis combinations $\{A_{1}B_{1}C_{1}\}$ and $\{A_{2}B_{1}C_{2}\}$ (see Appendix \ref{Appendix B}). Therefore, the QBER $Q_1$ caused by the white noise model is
\begin{eqnarray}\label{QBERQ1}
Q_1=Q_{111}=Q_{212}=\frac{1}{2}\left(1-F\right).
\end{eqnarray}

Then, we analyze the influence from the photon loss. In addition to the successful detection results $+1$ ($|+_x\rangle$ and $|+_y\rangle$) and $-1$ ($|-_x\rangle$ and $|-_y\rangle$), each user randomly defines the no-click event as an output $+1$ or $-1$. The violation of Bell-type inequalities can still be observed only if the photon loss is sufficiently weak (\emph{i.e.}, the global detection efficiency is sufficiently high). The QBER $Q_2$ caused by the photon loss is
\begin{eqnarray}\label{QBERQ2}
Q_2&=&\frac{3}{2}\eta^2(1-\eta)+\frac{3}{2}\eta(1-\eta)^2+\frac{1}{2}(1-\eta)^3 \nonumber\\
&=&\frac{1}{2}\left(1-\eta^3\right).
\end{eqnarray}

Thus, combining the decoherence and photon loss factors, the total QBER $\delta$ after transmitting photons through the noisy channel is
\begin{eqnarray}\label{totalQBER}
\delta&=&Q_1+Q_2=\frac{1-F}{2}\eta^3+\frac{1}{2}\left(1-\eta^3\right)\nonumber\\
&=&\frac{1}{2}-\frac{1}{2}\eta^3F.
\end{eqnarray}

In this way, the raw key irrelevance for the three users $H(A_{1}|B_{1},C_{1})$ and $H(A_{2}|B_{1},C_{2})$ have the form of
\begin{eqnarray}\label{irrelevance1}
H(A_{1}|B_{1},C_{1})=h(Q_{111})=h(\delta),\nonumber\\
H(A_{2}|B_{1},C_{2})=h(Q_{212})=h(\delta),
\end{eqnarray}
and the asymptotic key generation rate $r_\infty$ in Eq.~\eqref{rinfty2} can be calculated as
\begin{eqnarray}\label{rinfty3}
r_\infty=\left(p^2+\bar{p}^2\right)\left[\lambda H\left(A_1|E\right)+\bar{\lambda}H\left(A_2|E\right)-h\left(\delta\right)\right].
\end{eqnarray}

Based on above calculations, the theoretical value of the CHSH polynomial $S$ between Alice's and Bob's measurement results is
\begin{eqnarray}\label{noiseS}
S=2\sqrt{2}F\eta^3=2\sqrt{2}(1-2\delta).
\end{eqnarray}

By substituting Eq.~\eqref{keysecrecyrate4} and Eq.~\eqref{noiseS} into Eq.~\eqref{rinfty3}, the lower bound of DI QSS's key generation rate $r_\infty$ is given by
\begin{eqnarray}\label{rinfty4}
r_\infty\ge\left(p^2+\bar{p}^2\right)\left[g\left(\tilde{E} _\lambda\left(2\sqrt{2}\left(1-2\delta\right)\right)\right)-h(\delta)\right].
\end{eqnarray}

\begin{figure}
\includegraphics[scale=0.36]{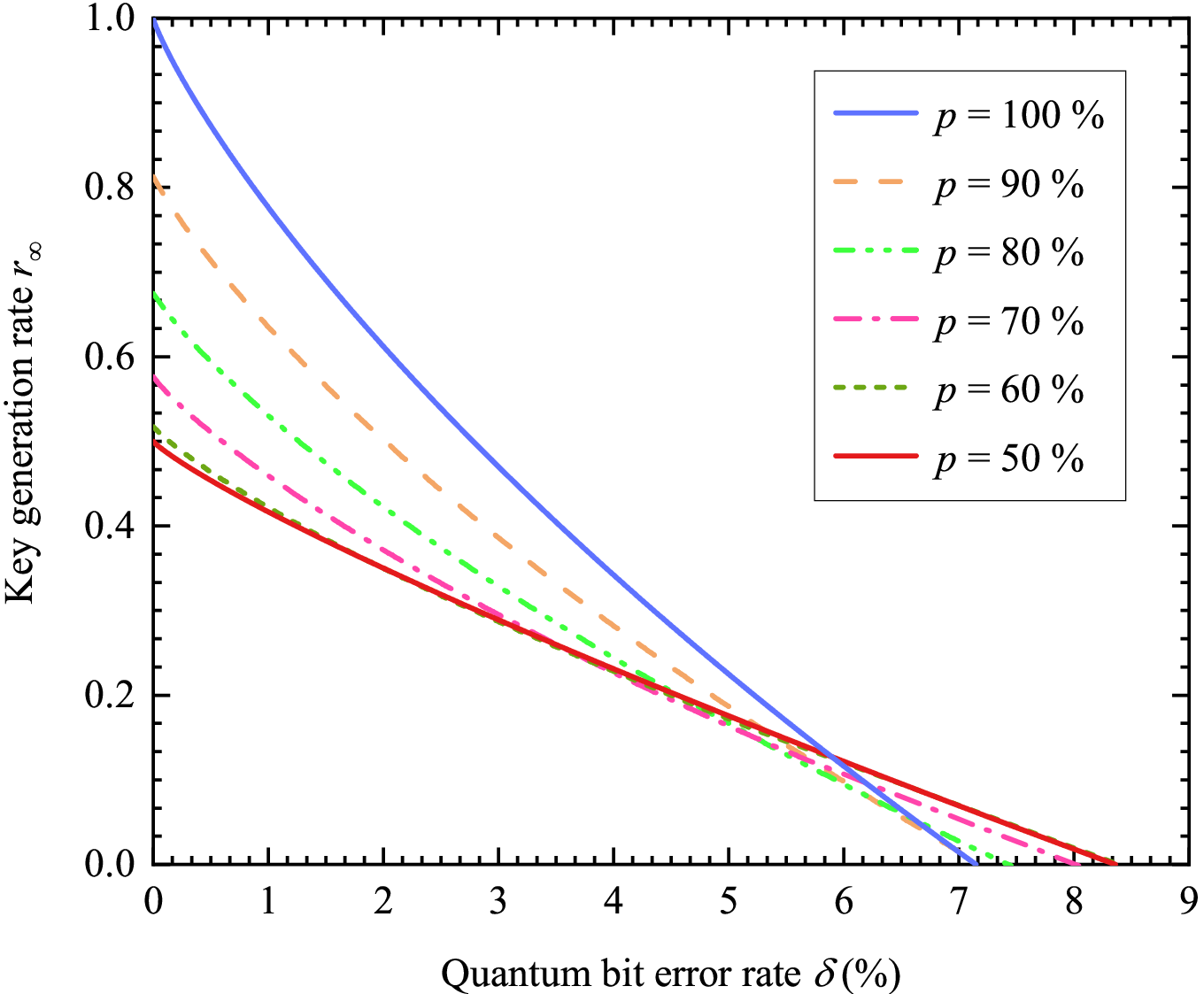}
\caption{(Color online) The key generation rate $r_\infty$ as a function of QBER. Alice (Charlie) chose $A_1$ ($C_1$) to measure the key generation rate with probabilities $p=100\%$, 90\%, 80\%, 70\%, 60\%, 50\%, respectively.}
\label{fig3}
\end{figure}

Fig.~\ref{fig3} demonstrates the key generation rate $r_\infty$ when Alice (Charlie) chooses $A_1$ ($C_1$) to generate the key with the probability $p=100\%, 90\%, 80\%, 70\%, 60\%, 50\%$, respectively. The case that Alice (Charlie) only uses $A_1$ ($C_1$) ($p=100\%$) to generate the raw key \cite{DIQSS1} has the noise tolerance threshold of 7.147\%. With the reduction of $p$, the noise tolerance threshold of the DI QSS protocol can be effectively increased, for Eve's uncertainty about the key generation basis is increased. The maximal noise tolerance threshold of 8.358\% can be achieved under the random key generation basis strategy ($p=50\%$). Such gain leads to a slight reduction in the maximal key generation rate due to the rise of the probability of measurement basis combination mismatch ($2p(1-p)$). For example, with $\delta=3\%$, the adoption of the random key generation basis strategy ($p=50\%$) reduces DI QSS's key generation rates from 0.470 to 0.289. The reduction effect reduces with the growth of noise error rate. When $\delta=5\%$, the adoption of the random key generation basis strategy reduces the key generation rate from 0.225 to 0.176.

\section{The DI QSS protocol with advanced random key generation basis strategy}\label{Section5}
For further increasing the noise tolerance threshold of our DI QSS protocol, we adopt the noise preprocessing strategy in our DI QSS protocol. Noise preprocessing is an important active improvement strategy, which can effectively increase the noise tolerance threshold of DI QKD \cite{DIQKD9,DIQKD12,DIQKD13} and DI QSS protocols \cite{DIQSS1}. Here, we combine the noise preprocessing with the random key generation basis strategies, which is called the advanced random key generation basis strategy.

The DI QSS protocol only requires minor changes as follows. When the measurement basis combination is $\{A_{1}B_{1}C_{1}\}$ or $\{A_{2}B_{1}C_{2}\}$, Alice flips her measurement result ($+1\rightarrow-1$, $-1\rightarrow+1$) with the probability of $q$. This operation essentially introduces some artificial noise to Alice's measurement results, which can further decrease the correlation among three users' measurement results, and increase Eve's uncertainty about  the key bits. As a result, the net effect on the key generation rate can be positive. After all the measurements, Alice will announce the flip probability $q$ in the error correction process. Then, three users apply a hash function to the raw key  bits to get the final secure key bits.

With the advanced random key generation basis strategy, the total noise qubit error rate (NQBER) $\delta_q$ consists of two components. First, the initial measurement results do not suffer from bit-flip errors but are flipped by Alice with the probability of $q$. Second, the initial measurement results suffer from bit-flip error and are not flipped by Alice with the probability of $\bar{q}=1-q$. Therefore, $\delta_q$ is given by
\begin{eqnarray}\label{totalQBER2}
\delta_q=q(1-\delta)+(1-q)\delta=q+(1-2q)\delta.
\end{eqnarray}

In this way, the tripartite raw key irrelevance becomes
\begin{eqnarray}\label{irrelevance2}
H(A_{1}|B_{1},C_{1})_q=H(A_{2}|B_{1},C_{2})_q=h(\delta_q).
\end{eqnarray}

According to the derivation of the DI QSS protocol with noise preprocessing strategy \cite{DIQSS1}, after the noise preprocessing, the lower bound of the total key secrecy to Eve is
\begin{eqnarray}\label{qkeysecrecyrate1}
&&H\left(A^q|E\right)=\lambda H\left(A_1^q|E\right)+\bar{\lambda}H\left(A_2^q|E\right)\nonumber\\
&&\geq g\left(\sqrt{\lambda\left\langle\bar{A_1}\otimes B\right\rangle^2+\bar{\lambda}\left\langle\bar{A_2}\otimes B'\right\rangle^2},q\right),
\end{eqnarray}
where
\begin{eqnarray}\label{gxq}
g(x,q)=1+\phi\left(\sqrt{\left(1-2q\right)^2+4q\left(1-q\right)x^2}\right)-\phi\left(x\right).\nonumber\\
\end{eqnarray}
The right part of Eq.~\eqref{qkeysecrecyrate1} has the same bound of that in Eq.~\eqref{keysecrecyrate3} \cite{DIQKD12} (see Appendix \ref{Appendix A}).

Finally, by substituting Eq.~\eqref{noiseS}, Eq.~\eqref{totalQBER2}-Eq.~\eqref{qkeysecrecyrate1} into Eq.~\eqref{rinfty2}, we can obtain a lower bound of the key generation rate $r_\infty^q$ of DI QSS with the advanced random key generation basis strategy as
\begin{eqnarray}\label{qrinfty1}
r_\infty^q &\ge& \left(p^2+\bar{p}^2\right)\Big[g\left(\tilde{E}_\lambda\left(2\sqrt{2}\left(1-2\delta\right)\right),q\right)\nonumber\\
&-& h\left(q+\left(1-2q\right)\delta\right)\Big].
\end{eqnarray}
When $q=0$, Eq.~\eqref{qrinfty1} degenerates into Eq.~\eqref{rinfty4}.

\begin{figure}
\includegraphics[scale=0.37]{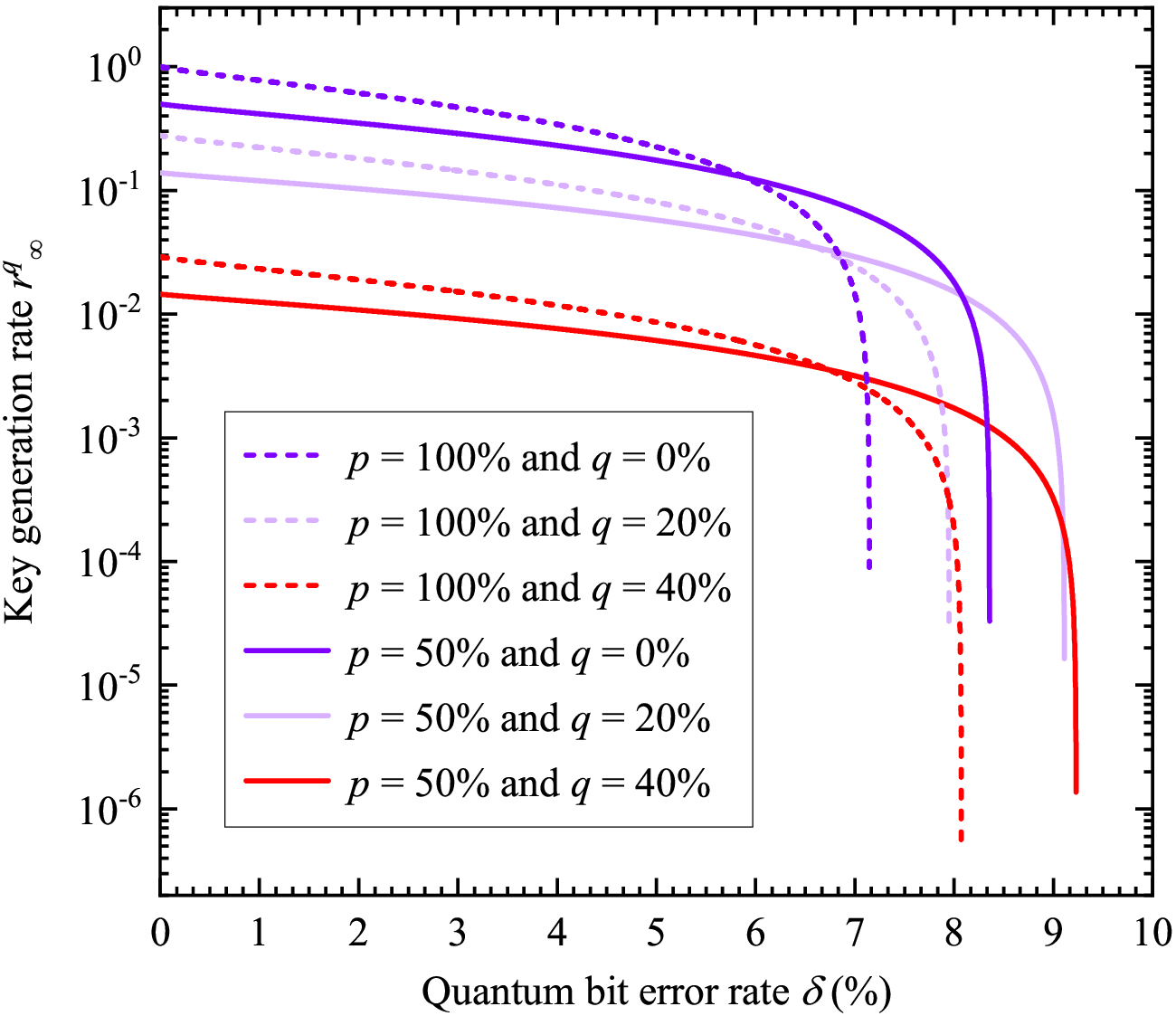}
\caption{(Color online) The key generation rate $r_\infty^q$ as a function of the QBER $\delta_q$ with different values of $p$ and $q$. The maximal noise tolerance threshold is further improved by using the noise preprocessing strategy ($q=20\%$, 40\%).}
\label{fig4}
\end{figure}

In Fig.~\ref{fig4}, the noise tolerance thresholds of the DI QSS protocol with one key generation basis ($p=100\%$, dashed lines) and two random key generation basis ($p=50\%$, solid lines) both increase with the growth of the flipping probability $q$. When using the same level (same color) of artificial noise (same value of $q$), the noise tolerance thresholds of the DI QSS protocol with the advanced random key generation basis strategy ($p=50\%$, solid lines) are all higher than that only uses one key generation basis ($p=100\%$, dashed lines). In detail, the advanced random key generation basis strategy ($q=40\%$ and $p=50\%$) can improve the noise tolerance of the DI QSS protocol from 7.148\% (blue dashed line) to 9.231\% (red solid line) (29.16\% growth).

\begin{figure}
\includegraphics[scale=0.37]{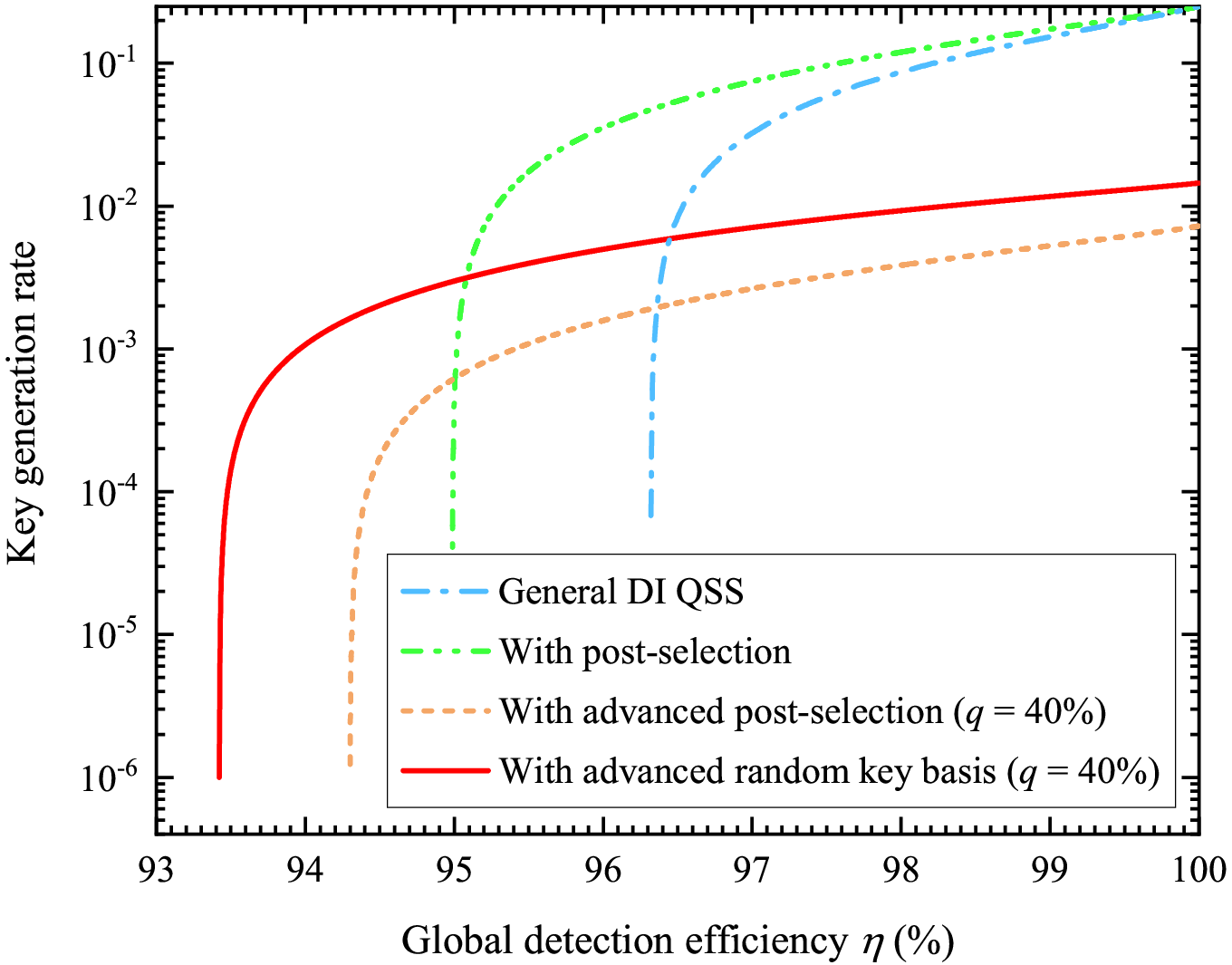}
\caption{(Color online) The key generation rate of DI QSS in the case of using the four active improvement strategies as a function of the global detection efficiency $\eta$ with the fidelity $F=1$.}
\label{fig5}
\end{figure}

We define the global detection efficiency $\eta=\eta_t\eta_d\eta_c$, where $\eta_d$ is the detection efficiency of the photon detector and $\eta_c$ is the coupling efficiency of the photon to the fiber. The photon transmission efficiency in noisy quantum channel is $\eta_t=10^{-\frac{\alpha L}{10}}$, where $L$ denotes the photon transmission distance, $\alpha=0.2$ dB/km for the standard optical fiber. In theory, the growth of noise tolerance threshold would effectively reduce the global detection efficiency threshold. Then, we calculate the global detection efficiency threshold of the DI QSS protocol with the fidelity $F=1$. From Fig.~\ref{fig5}, the adoption of the advanced random key generation basis strategy ($q=40\%$) can reduce the global detection efficiency threshold from initial 96.32\% to 93.42\%, which is further lower than 94.30\% corresponding to the adoption of the advanced postselection strategy (combing noise preprocessing with $q=40\%$ and the postselection strategies) \cite{DIQSS1}. As a result, the advanced random key generation basis strategy can effectively reduce the requirement on experimental equipment's performance and facilitate DI QSS's experimental realization.

\begin{figure}
\includegraphics[scale=0.36]{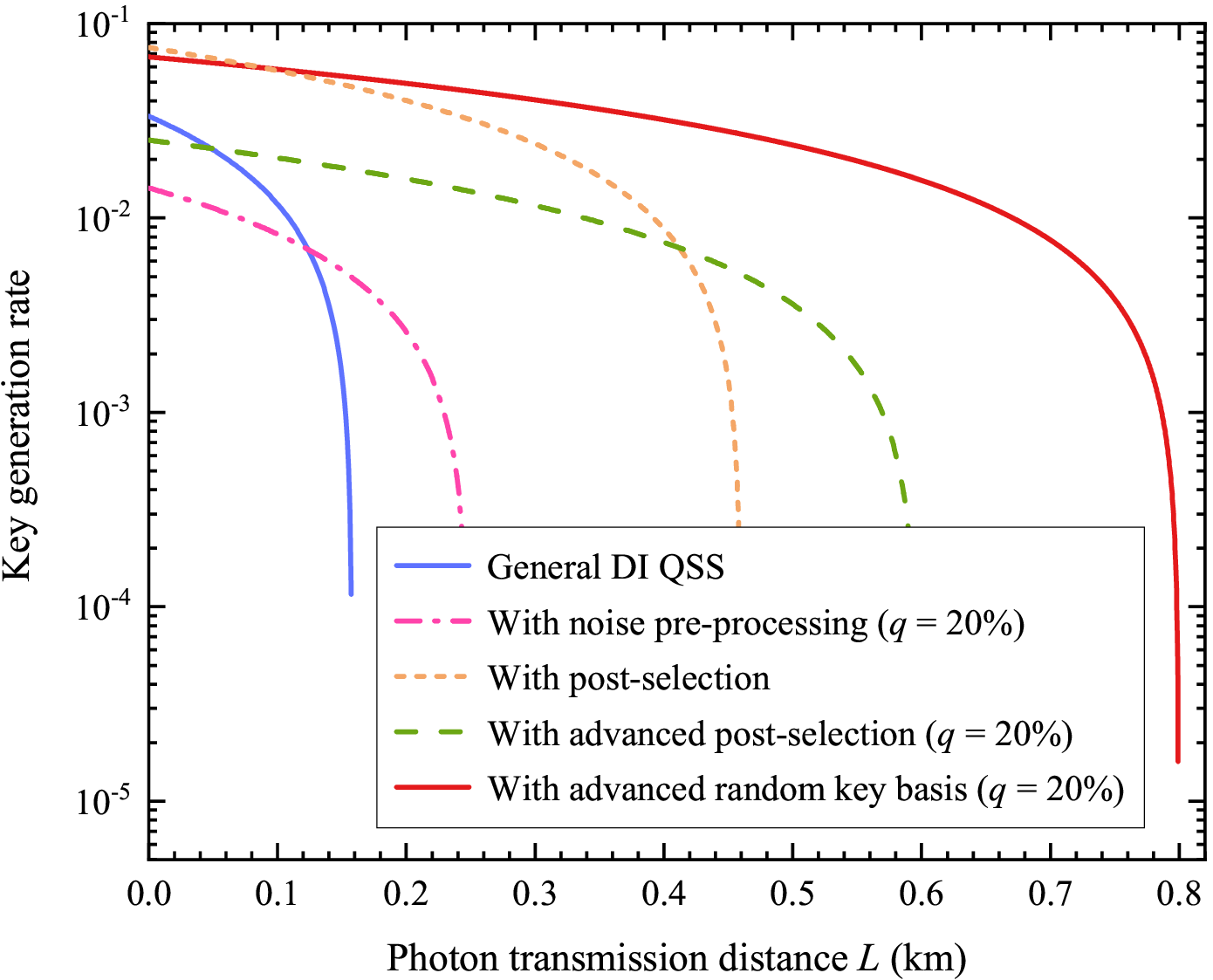}
\caption{(Color online) The key generation rate of the DI QSS protocol in five cases as a function of the photon transmission distance $L$. Here, we control the efficiency $\eta_d=98\%$, $\eta_c=99\%$ and the fidelity $F=1$.}
\label{fig6}
\end{figure}

Fig.~\ref{fig6} simulates the key generation rate of the DI QSS protocol in five scenarios altered with the photon transmission distance $L$. The five scenarios include the DI QSS without any active improve strategy, with noise preprocessing strategy($q=20\%$), with postselection strategy, with advanced post selection strategy (combining noise preprocessing and post selection strategies), and advanced random key generation basis strategy. Recently, the superconducting nanowire single-photon detector (SNSPD) with a detection efficiency of 98\% at 1550 nm band was reported \cite{SNSD}. Therefore, we assume $\eta_d=98\%$ and $\eta_c=99\%$ in our simulations. The DI QSS protocol with the advanced random key generation basis strategy ($q=20\%$) has the longest photon transmission distance of 0.799 km. This means that the maximal secure communication distance between any two users achieves 1.384 km. If we increase the value of $q$ to 40\%, the maximal secure communication distance between any two users can achieve 1.43 km, which is about 5.5 times of the initial value (0.26 km) \cite{DIQSS1}. Meanwhile, DI QSS with the advanced random key generation basis strategy ($q=20\%$) has the highest key generation rate when photons are transmitted over long distances. For example, in the case of $L=0.3$ km, its key generation rate is about 3.5 times of that with the advanced postselection strategy ($q=20\%$). Based on the comparison results, the DI QSS protocol with advanced random key generation basis strategy has big advantages in practical applications.

\section{Discussion and conclusion}\label{Section6}
In the paper, we propose the DI QSS protocol with the advanced random key generation basis, which can effectively improve DI QSS's noise tolerance and maximal secure photon transmission distance. In this work, we analyze DI QSS's conditional entropy bound by simplifying it to the qubit bound, and numerically and analytically evaluate its key generation rate. The advanced random key generation basis strategy can effectively increase Eve's uncertainty about the key generation basis and the measurement results, which eventually improves the protocol's resistance to noise.

Here, we consider the tripartite DI QSS protocol. Actually, we can use the $m$-party Svetlichny polynomial with the form of CHSH polynomials to extend it to the general multipartite DI QSS protocol with advanced random key generation basis strategy \cite{nonlocal4}. The $m$-party Svetlichny inequality has the form of
\begin{eqnarray}\label{mSvetlichny}
S_{m}=\langle S_{m-1}M_{2}\rangle+\langle S_{m-1}'M_{1}\rangle\leq 2^{m-1}.
\end{eqnarray}
We can use the violation of $m$-party Svetlichny inequality to guarantee the nonlocal correlation among the particles in $m$ users and realize the general $m$-party DI QSS with advanced random key generation basis strategy.

In our DI QSS protocol, Alice, Bob, and Charlie all independently select the measurement basis to measure their photons, which results in the final key generation rate decreasing due to the ``sifting'' effect (only two of the twelve basis combinations are used to generate keys, the
measurement results corresponding to the basis combinations $\{A_1B_1C_2\}$ and $\{A_2B_1C_1\}$ should be discarded). In particularly, with the growth of the user or measurement settings number, the ``sifting'' effect would be more serious. Two possible methods may be applied to avoid this degradation \cite{DIQKD11}. The first method is to use the quantum memory. In detail, three users receive and store all quantum states in quantum memories. Then, they publicly announce their basis choices and measure their own photon after agreeing on the basis choices (avoiding the case of choosing $\{A_1B_1C_2\}$ and $\{A_2B_1C_1\})$. The second method is the most hopeful one, where Alice, Bob, and Charlie need to share a long enough pre-shared key in advance to coordinate their measurement basis selection during the key generation rounds.

Finally, we discuss the experimental realization of our DI QSS protocol. The DI QSS protocol requires to distribute high-quality GHZ state through quantum channels. The generation of three-photon GHZ states with high fidelity has been extensively studied \cite{GHZstate1,GHZstate2,GHZstate3,GHZstate4}. One common GHZ state generation method depends on the post selection \cite{GHZstate1,GHZstate3}, in which the generated GHZ state is destroyed by the photon detection and cannot be used. In 2014, Hamel \emph{et al.} \cite{GHZstate2} eliminated the limitation of the outcome post selection by cascading two SPDC sources for directly generating three-photons polarization GHZ states with the fidelity of 86\%. Recently, by using the phase-stable source, the fidelity of the target GHZ state was increased to over 96\% \cite{GHZstate4}. Only the phase-flip error may occur in the practical GHZ state generation \cite{GHZstate2,GHZstate4}, which will cause the quantum bit error in the key generation process. In this way, the imperfect GHZ state generated in Ref.~\cite{GHZstate4} would cause the QBER of 4\%. This QBER is below the noise threshold 9.231\%, so that the current GHZ state generation technology in Ref.~\cite{GHZstate4} satisfies the requirement of our DI QSS protocol with advanced random key generation basis strategy. Combining the practical high-efficient SNSPD, our DI QSS protocol with advanced random key generation basis strategy has potential to demonstrate with current experimental technology. On the other hand, the photon transmission loss and decoherence caused by the practical noisy channels severely degrade the entanglement and reduce DI QSS's secret key generation rate. Since DI QSS has to transmit entanglement photon pairs to multiple users, any photon loss that occurs during transmission can break the entanglement, so it has relatively low resistance to photon loss. Quantum repeater provides a promising way to construct high-quality long-distance entanglement channel \cite{repeater1,repeater2}. Combining quantum repeater with the DI QSS protocol may further increase the secure communication distance and realize the long-distance DI QSS.

In conclusion, DI QSS can provide the highest security for QSS under practical imperfect experimental conditions. However, DI QSS protocol has quite low noise tolerance threshold and high global detection efficiency requirement, which largely limit its experimental demonstration and application. To reduce DI QSS's experimental requirements and improve its performance in practical noisy experiment, we propose a new DI QSS protocol with advanced random key generation basis strategy. The advanced random key generation basis strategy combine the random key generation basis, noise preprocessing, and postselection strategies, which can effectively increase the total key secrecy rate to Eve, and thus effectively increase DI QSS's noise tolerance threshold from 7.147\% to 9.231\% (29.16\% growth) and reduce the global detection efficiency requirement from initial 96.32\% to 93.41\% ($q\rightarrow 50\%$). In this way, the maximal secure communication distance between any two users can be increased from 0.26 km to 1.43 km (about 5.5 times). Meanwhile, our DI QSS can reuse some discarded items in original DI QSS protocol \cite{DIQSS1} and reduce the entanglement resource consumption. DI QSS requires to distribute high-quality GHZ state. The requirement for high-quality GHZ state can be satisfied with current GHZ source. Our DI QSS protocol can be extended to the general multipartite condition and has potential to demonstrate with current experimental technology, which can effectively promote DI QSS's experimental demonstration and applications in the future.

\section*{Acknowledgement}
This work was supported by the National Natural Science Foundation of China under Grants No. 12175106 and No. 92365110, the Natural Science Foundation of Jiangsu Province of China under Grants No. BK20240611 and No. BK20240612.

\appendix
\section{Reduce the DI conditional entropy bound of Eve to the qubit bound}\label{Appendix A}
\setcounter{equation}{0}
\setcounter{subsection}{0}
\renewcommand{\theequation}{A.\arabic{equation}}

In Sec.~\ref{Section4.1}, we first constrain the conditional entropy of Eve's knowledge about Alice's key with the correlation $\left\langle\bar{A_i}\otimes B\right\rangle$. Then, we establish a full DI qubit bound on $\left\langle\bar{A_i}\otimes B\right\rangle$. Finally, we perform a convexity analysis of the DI conditional entropy bound \cite{DIQKD12}.

In the DI scenario, we note that all devices are untrusted, so that the measurement basis used by three users may be potentially under Eve's control. As a result, all measurement bases need to be reparametrized in the following analysis. By simplifying the DI conditional entropy bound to the qubit system, the two-basis (random key generation basis) variant has a conditional entropy bound as
\begin{eqnarray}\label{Akeysecrecyrate3}
H\left(A|E\right)&=&\lambda H\left(A_1|E\right)+\bar{\lambda}H\left(A_2|E\right)\nonumber\\
&\geq & g\left(\sqrt{\lambda\left\langle\bar{A_1}\otimes B\right\rangle^2+\bar{\lambda}\left\langle\bar{A_2}\otimes B'\right\rangle^2}\right),\nonumber\\
\end{eqnarray}
where $\bar{A_1}$ and $\bar{A_2}$ are the Pauli observations orthogonal to $A_1$ and $A_2$ in the X-Y plane of the Bloch sphere, respectively. $B$ and $B'$ are observations with any given eigenvalue $\pm 1$ on Bob's subsystem. In DI-type protocols, we do not directly obtain the correlation between Alice's and Bob's subsystems ($\left\langle\bar{A_i}\otimes B\right\rangle$, $i=1,2$). We maximally estimate Eve's information about the measurement results of $A_1$ and $A_2$ by the correlation operator of Bob's subsystem with the virtual complementary $\bar{A_1}$ and $\bar{A_2}$. Let the lower bound function $g(x)=1-h(\frac{1}{2}+\frac{1}{2}x)$.

In terms of $H(A|E)$, in order to find the optimal value of the conditional entropy bound, we need to parameterize all the qubit operators in the X-Y plane of the Bloch sphere. Let the observations $B=\cos \varphi _B\sigma_x+\sin \varphi _B\sigma_y$ and $B'=\cos \varphi _B'\sigma_x+\sin \varphi _B'\sigma_y$. The $g(\sqrt{x})$ function is monotonically increasing and convex in the $x\in[0,1]$. Therefore, we choose the optimal $\varphi_B$ and $\varphi_B'$ by Cauchy-Schwarz inequality to maximize the lower bound of $H(A|E)$ with the form of
\begin{widetext}
\begin{eqnarray}\label{qubit}
\lambda\left\langle\bar{A_1}\otimes B\right\rangle^2&+&\bar{\lambda}\left\langle\bar{A_2}\otimes B'\right\rangle^2\nonumber\\
&=&\lambda\left\langle\bar{A_1}\otimes \left(\cos \varphi _B\sigma_x+\sin \varphi _B\sigma_y\right)\right\rangle^2+\bar{\lambda}\left\langle\bar{A_2}\otimes \left(\cos \varphi _B'\sigma_x+\sin \varphi _B'\sigma_y\right)\right\rangle^2\nonumber\\
&=&\lambda \left(\cos \varphi_B\left\langle\bar{A_1}\otimes \sigma_x\right\rangle+\sin \varphi_B\left\langle\bar{A_1}\otimes \sigma_y\right\rangle \right)^2+\bar{\lambda} \left(\cos \varphi_B'\left\langle\bar{A_2}\otimes \sigma_x\right\rangle+\sin \varphi_B'\left\langle\bar{A_2}\otimes \sigma_y\right\rangle \right)^2\nonumber\\
&\le& \lambda \left[\sqrt{\left(\cos^2 \varphi_B+\sin^2 \varphi_B\right)\left(\left\langle\bar{A_1}\otimes \sigma_x\right\rangle^2+\left\langle\bar{A_1}\otimes \sigma_y\right\rangle^2\right)} \right]^2\nonumber\\
&+&\bar{\lambda} \left[\sqrt{\left(\cos^2 \varphi_B'+\sin^2 \varphi_B'\right)\left(\left\langle\bar{A_2}\otimes \sigma_x\right\rangle^2+\left\langle\bar{A_2}\otimes \sigma_y\right\rangle^2\right)} \right]^2\nonumber\\
&=&\lambda\left(\left\langle\bar{A_1}\otimes \sigma_x\right\rangle^2+\left\langle\bar{A_1}\otimes \sigma_y\right\rangle^2\right)
+\bar{\lambda}\left(\left\langle\bar{A_2}\otimes \sigma_x\right\rangle^2+\left\langle\bar{A_2}\otimes \sigma_y\right\rangle^2\right).
\end{eqnarray}
\end{widetext}

To maximize Eve's uncertainty about the measurement basis, we first parameterize Alice's measurement bases as
\begin{eqnarray}\label{Alices basis}
A_1&=&\cos\frac{\varphi_A}{2}\sigma_x-\sin\frac{\varphi_A}{2}\sigma_y,\nonumber\\
A_2&=&\cos\frac{\varphi_A}{2}\sigma_x+\sin\frac{\varphi_A}{2}\sigma_y.
\end{eqnarray}
Their corresponding virtual complementary operators are
\begin{eqnarray}\label{Alices basis2}
\bar{A_1}&=&\sin\frac{\varphi_A}{2}\sigma_x+\cos\frac{\varphi_A}{2}\sigma_y,\nonumber\\
\bar{A_2}&=&-\sin\frac{\varphi_A}{2}\sigma_x+\cos\frac{\varphi_A}{2}\sigma_y.
\end{eqnarray}

Substituting $\bar{A_1}$ and $\bar{A_2}$ into Eq.~\eqref{qubit} can obtain the objective function of the optimization problem as
\begin{widetext}
\begin{eqnarray}\label{qubit2}
\lambda\left\langle\bar{A_1}\otimes B\right\rangle^2&+&\bar{\lambda}\left\langle\bar{A_2}\otimes B'\right\rangle^2 \le \lambda\left(\left\langle\bar{A_1}\otimes \sigma_x\right\rangle^2+\left\langle\bar{A_1}\otimes \sigma_y\right\rangle^2\right)+\bar{\lambda}\left(\left\langle\bar{A_2}\otimes \sigma_x\right\rangle^2+\left\langle\bar{A_2}\otimes \sigma_y\right\rangle^2\right)\nonumber\\
&=&\lambda\left[\left\langle\left(\sin\frac{\varphi_A}{2}\sigma_x+\cos\frac{\varphi_A}{2}\sigma_y\right)\otimes \sigma_x\right\rangle^2+\left\langle\left(\sin\frac{\varphi_A}{2}\sigma_x+\cos\frac{\varphi_A}{2}\sigma_y\right)\otimes \sigma_y\right\rangle^2\right]\nonumber\\
&+&\bar{\lambda}\left[\left\langle\left(-\sin\frac{\varphi_A}{2}\sigma_x+\cos\frac{\varphi_A}{2}\sigma_y\right)\otimes \sigma_x\right\rangle^2+\left\langle\left(-\sin\frac{\varphi_A}{2}\sigma_x+\cos\frac{\varphi_A}{2}\sigma_y\right)\otimes \sigma_y\right\rangle^2\right]\nonumber\\
&=&\sin^2\frac{\varphi_A}{2}\left(E_{XX}^2+E_{XY}^2\right)+\cos^2\frac{\varphi_A}{2}\left(E_{YX}^2+E_{YY}^2\right)\nonumber\\
&+&2\left(2\lambda-1\right)\sin\frac{\varphi_A}{2}\cos\frac{\varphi_A}{2}\left(E_{XX}E_{YX}+E_{XY}E_{YY}\right),
\end{eqnarray}
\end{widetext}
where Pauli operators $E_{XX}=\left\langle\sigma_x\otimes\sigma_x\right\rangle$, $E_{YX}=\left\langle\sigma_y\otimes\sigma_x\right\rangle$, $E_{XY}=\left\langle\sigma_x\otimes\sigma_y\right\rangle$, and $E_{YY}=\left\langle\sigma_y\otimes\sigma_y\right\rangle$.

Now, we need to find constraints on the optimization problem \eqref{qubit2} in terms of the nonlocal correlations $S$ and the Pauli operators ($E_{XX}$, $E_{YX}$, $E_{XY}$, and $E_{YY}$) to further bound \eqref{Akeysecrecyrate3}.

Firstly, in terms of DI QSS, we emphasize that the tripartite nonlocal correlation is embodied in the CHSH expectation values $S$ between Alice and Bob in Sec.~\ref{Section2.2}. In imperfect measurements, maximizing the CHSH expectation value $S$ between Alice and Bob can increase the key secrecy rate to Eve. For the first case of the CHSH polynomial $S$ between Alice and Bob ($S=S_{AB}$), we rewrite $S_{AB}$ in Eq.~\eqref{CHSH} into qubit operator form
\begin{widetext}
\begin{eqnarray}\label{AreduceCHSH1}
S_{AB}&=&\left\langle A_1\otimes B_2\right\rangle+\left\langle A_2\otimes B_2\right\rangle+\left\langle A_1\otimes B_3\right\rangle-\left\langle A_2\otimes B_3\right\rangle \nonumber\\
&=&\left\langle\left(A_1+A_2\right)\otimes B_2\right\rangle+\left\langle\left(A_1-A_2\right)\otimes B_3\right\rangle\nonumber\\
&=&2\cos\frac{\varphi_A}{2}\left\langle\sigma_x\otimes B_2\right\rangle-2\sin\frac{\varphi_A}{2}\left\langle\sigma_y\otimes B_3\right\rangle.
\end{eqnarray}
\end{widetext}

We use $B_2$ and $B_3$ to maximize Eq.~\eqref{AreduceCHSH1} by the Cauchy-Schwarz inequality as
\begin{widetext}
\begin{eqnarray}\label{AreduceCHSH12}
\frac{S_{AB}}{2}&=&\cos\frac{\varphi_A}{2}\left\langle\sigma_x\otimes B_2\right\rangle-\sin\frac{\varphi_A}{2}\left\langle\sigma_y\otimes B_3\right\rangle\nonumber\\
&=&\cos\frac{\varphi_A}{2}\left(\cos\varphi_{B_2}\left\langle\sigma_x\otimes\sigma_x\right\rangle+\sin\varphi_{B_2}\left\langle\sigma_x\otimes\sigma_y\right\rangle\right)\nonumber\\
&-&\sin\frac{\varphi_A}{2}\left(\cos\varphi_{B_3}\left\langle\sigma_y\otimes\sigma_x\right\rangle+\sin\varphi_{B_3}\left\langle\sigma_y\otimes\sigma_y\right\rangle\right) \nonumber\\
&\le&\cos\frac{\varphi_A}{2}\sqrt{\left\langle\sigma_x\otimes\sigma_x\right\rangle^2+\left\langle\sigma_x\otimes\sigma_y\right\rangle^2}-\sin\frac{\varphi_A}{2}\sqrt{\left\langle\sigma_y\otimes\sigma_x\right\rangle^2+\left\langle\sigma_y\otimes\sigma_y\right\rangle^2}\nonumber\\
&=&\cos\frac{\varphi_A}{2}\sqrt{E_{XX}^2+E_{XY}^2}-\sin\frac{\varphi_A}{2}\sqrt{E_{YX}^2+E_{YY}^2}.
\end{eqnarray}
\end{widetext}

Similarly, for the second case of the CHSH polynomial $S$ between Alice and Bob ($S=S_{AB}'$), we rewrite $S_{AB}'$ in Eq.~\eqref{CHSH} with the form of qubit operator as
\begin{widetext}
\begin{eqnarray}\label{AreduceCHSH2}
S_{AB}'&=&\left\langle A_2\otimes B_3\right\rangle+\left\langle A_2\otimes B_2\right\rangle+\left\langle A_1\otimes B_3\right\rangle-\left\langle A_1\otimes B_2\right\rangle \nonumber\\
&=&\left\langle\left(A_1+A_2\right)\otimes B_3\right\rangle-\left\langle\left(A_1-A_2\right)\otimes B_2\right\rangle\nonumber\\
&=&2\cos\frac{\varphi_A}{2}\left\langle\sigma_x\otimes B_3\right\rangle+2\sin\frac{\varphi_A}{2}\left\langle\sigma_y\otimes B_2\right\rangle.
\end{eqnarray}
\end{widetext}

We use $B_2$ and $B_3$ to maximize Eq.~\eqref{AreduceCHSH2} by the Cauchy-Schwarz inequality as
\begin{widetext}
\begin{eqnarray}\label{AreduceCHSH22}
\frac{S_{AB}'}{2}&=&\cos\frac{\varphi_A}{2}\left\langle\sigma_x\otimes B_3\right\rangle+\sin\frac{\varphi_A}{2}\left\langle\sigma_y\otimes B_2\right\rangle \nonumber\\
&=&\cos\frac{\varphi_A}{2}\left(\cos\varphi_{B_3}\left\langle\sigma_x\otimes\sigma_x\right\rangle+\sin\varphi_{B_3}\left\langle\sigma_x\otimes\sigma_y\right\rangle\right)\nonumber\\
&+&\sin\frac{\varphi_A}{2}\left(\cos\varphi_{B_2}\left\langle\sigma_y\otimes\sigma_x\right\rangle+\sin\varphi_{B_2}\left\langle\sigma_y\otimes\sigma_y\right\rangle\right)\nonumber\\
&\le&\cos\frac{\varphi_A}{2}\sqrt{\left\langle\sigma_x\otimes\sigma_x\right\rangle^2+\left\langle\sigma_x\otimes\sigma_y\right\rangle^2}+\sin\frac{\varphi_A}{2}\sqrt{\left\langle\sigma_y\otimes\sigma_x\right\rangle^2+\left\langle\sigma_y\otimes\sigma_y\right\rangle^2}\nonumber\\
&=&\cos\frac{\varphi_A}{2}\sqrt{E_{XX}^2+E_{XY}^2}+\sin\frac{\varphi_A}{2}\sqrt{E_{YX}^2+E_{YY}^2}.
\end{eqnarray}
\end{widetext}

In conclusion, based on Eq.~\eqref{Bell} and the absolute value inequality $|a\pm b|\leq|a|+|b|$, we have
\begin{eqnarray}\label{ACHSH1}
\frac{S}{2}&\le& \left|\cos\frac{\varphi_A}{2}\right|\sqrt{E_{XX}^2+E_{XY}^2}\nonumber\\
&+&\left|\sin\frac{\varphi_A}{2}\right|\sqrt{E_{YX}^2+E_{YY}^2}.
\end{eqnarray}

Then, we bound $E_{XX}$, $E_{YX}$, $E_{XY}$, $E_{YY}$. For arbitrary normalized Bloch vectors $\textbf{a}=(a_x\quad a_y)$ and $\textbf{b}=(b_x\quad b_y)$ in the X-Y plane, the linear combinations $\textbf{a} \cdot \boldsymbol{\sigma}$ and $\textbf{b} \cdot \boldsymbol{\sigma}$ have eigenvalues $\pm 1$, where $\boldsymbol{\sigma}=(\sigma_x\quad \sigma_y\quad \sigma_z$). Thus, for any measurement operator, we have
\begin{eqnarray}\label{qubit3}
\left\langle\left(\textbf{a}\cdot\boldsymbol{\sigma}\right)\otimes\left(\textbf{b}\cdot\boldsymbol{\sigma}\right)\right\rangle&=&\left\langle\left(a_x\sigma_x+a_y\sigma_y\right)\otimes\left(b_x\sigma_x+b_y\sigma_y\right)\right\rangle \nonumber\\
&=&\sum_{m,n}a_mb_n\left\langle\sigma_m\otimes\sigma_n\right\rangle\le 1.
\end{eqnarray}
where $m,n\in\{x,y\}$. The right part of Eq.~\eqref{qubit3} can be given in terms of matrix as
\begin{eqnarray}\label{matrix}
&&\sum_{m,n}a_mb_n\left\langle\sigma_m\otimes\sigma_n\right\rangle \nonumber\\
&&=
\left(\begin{matrix}
 a_x\quad a_y
\end{matrix} \right)
\begin{pmatrix}
 \left\langle\sigma_x\otimes\sigma_x\right\rangle  & \left\langle\sigma_x\otimes\sigma_y\right\rangle\\
 \left\langle\sigma_y\otimes\sigma_x\right\rangle  & \left\langle\sigma_y\otimes\sigma_y\right\rangle
\end{pmatrix}
\binom{b_x}{b_y} \nonumber\\
&&= \textbf{a}^{\mathsf{T}}\textbf{Eb},
\end{eqnarray}
where,
\begin{eqnarray}\label{E}
\textbf{E}=
\begin{pmatrix}
 E_{XX} & E_{XY}\\
 E_{YX} & E_{YY}
\end{pmatrix}.
\end{eqnarray}

Since \textbf{a} and \textbf{b} are arbitrary Bloch vectors in the X-Y plane, maximizing Eq.~\eqref{matrix} by $\textbf{a}=\textbf{b}$ yields $\|\textbf{E}\|\le 1$, which is equivalent to $\textbf{EE}^{\mathsf{T}} \le I$. In matrix theory, this means that the matrix
\begin{widetext}
\begin{eqnarray}\label{IEE}
I-\textbf{EE}^{\mathsf{T}}=
\begin{pmatrix}
 1-E_{XX}^2-E_{XY}^2 & -E_{XX}E_{YX}-E_{XY}E_{YY}\\
 -E_{XX}E_{YX}-E_{XY}E_{YY} & 1-E_{YX}^2-E_{YY}^2
\end{pmatrix}
\end{eqnarray}
\end{widetext}
is semi-positive definite. According to the Sylvester criterion, the matrix is semi-positive definite if and only if all principal minors of \eqref{IEE} are non-negative determinants. We derive the constraints for the matrix as
\begin{widetext}
\begin{eqnarray}\label{Abound}
\begin{cases}
 1-E_{XX}^2-E_{XY}^2\ge 0,\\
 1-E_{YX}^2-E_{YY}^2\ge 0,\\
 \left(1-E_{XX}^2-E_{XY}^2\right)\left(1-E_{YX}^2-E_{YY}^2\right)-\left(E_{XX}E_{YX}+E_{XY}E_{YY}\right)^2\ge 0.
\end{cases}
\end{eqnarray}
\end{widetext}

To constrain the objective function Eq.~\eqref{qubit2}, we denote $\lambda\left\langle\bar{A_1}\otimes B\right\rangle^2+\bar{\lambda}\left\langle\bar{A_2}\otimes B'\right\rangle^2\ge E_\lambda(S)^2$ and introduce new variables to simplify the analysis of the maximization problem. Let
\begin{eqnarray}\label{variable}
&&s=\sin\frac{\varphi_A}{2},~~c=\cos\frac{\varphi_A}{2},~~\varDelta=\cos(\gamma-\mu),\nonumber\\
&&E_{XX}=g\cos\gamma,~~E_{XY}=g\sin\gamma,\nonumber\\
&&E_{YX}=h\cos\mu,~~E_{YY}=h\sin\mu.
\end{eqnarray}

Based on the constraints \eqref{Abound}, the maximization problem \eqref{qubit2} is rewritten as
\begin{eqnarray}\label{Amaxproblem}
E_\lambda(S)^2=\max& s^2g^2+c^2h^2+2(2\lambda-1)scgh\varDelta. \nonumber\\
s.t.& cg+sh\geq S/2 \nonumber\\
& g^2 \leq 1 \nonumber\\
& h^2 \leq 1 \nonumber\\
& (1-g^2)(1-h^2)\geq g^2h^2\varDelta^2 \nonumber\\
& c^2+s^2=1 \nonumber\\
& \varDelta ^{2} \leq 1
\end{eqnarray}

Eq.~\eqref{Amaxproblem} can be reduced to a series of semidefinite program problems by using the Lasserre hierarchy \cite{Lasserre1,Lasserre2}, which gives the optimal entropy bound in a reasonable time using the MOSEK solver \cite{mosek}. When $\lambda=\frac{1}{2}$ (\emph{i.e.}, $p=\frac{1}{2}$), Eq.~\eqref{Amaxproblem} degenerates into a quadratic root-finding problem
\begin{eqnarray}\label{root-finding}
4x(2-x)+2(S^2+2)+S(x-S)\sqrt{2(1+x)}=0.\nonumber\\
\end{eqnarray}
In the limit $2<S\le 2\sqrt{2}$, we can obtain a strict analytical solution. A reliable solution method for this problem can be found in Ref.~\cite{DIQKD12}.

Finally, we obtain DI QSS's conditional entropy bound in \eqref{Akeysecrecyrate3} as
\begin{eqnarray}\label{AHAiiE2}
H\left(A|E\right)\ge g\left(\tilde{E}_\lambda(S)\right),
\end{eqnarray}
where $\tilde{E}_\lambda(S)\equiv \sqrt{\tilde{E}_\lambda(S)^2}$. $\tilde{E}_\lambda(S)^2$ is the maximal value of $E_\lambda(S)^2$ by numerically and analytically solving \eqref{Amaxproblem}.
\section{The white noise model}\label{Appendix B}
\setcounter{equation}{0}
\setcounter{subsection}{0}
\renewcommand{\theequation}{B.\arabic{equation}}

Decoherence caused by channel noise can severely weaken the nonlocal correlation among users' measurement results. Here, we consider the white noise model, \emph{i.e.}, the target GHZ state may degenerate into eight possible GHZ states with equal probability. Alice, Bob, and Charlie share noisy GHZ states with the form of
\begin{eqnarray}\label{BGHZ}
\rho_{ABC}=F|GHZ\rangle\langle GHZ|+\frac{1-F}{8} I,
\end{eqnarray}
where the fidelity $F$ is the probability that no error occurs in the photon state, and the unit matrix $I$ is composed by the density matrices of the eight possible noisy GHZ states as
\begin{eqnarray}\label{BGHZ2}
I&=&|GHZ_1^+\rangle\langle GHZ_1^+|+|GHZ_1^-\rangle\langle GHZ_1^-| \nonumber\\
&+&|GHZ_2^+\rangle\langle GHZ_2^+|+|GHZ_2^-\rangle\langle GHZ_2^-|\nonumber\\
&+&|GHZ_3^+\rangle\langle GHZ_3^+|+|GHZ_3^-\rangle\langle GHZ_3^-|\nonumber\\
&+&|GHZ_4^+\rangle\langle GHZ_4^+|+|GHZ_4^-\rangle\langle GHZ_4^-|,
\end{eqnarray}
where
\begin{eqnarray}\label{BGHZ3}
|GHZ_1^\pm\rangle&=&\frac{1}{\sqrt{2}}\left(|HHH\rangle\pm |VVV\rangle\right),\nonumber\\
|GHZ_2^\pm\rangle&=&\frac{1}{\sqrt{2}}\left(|HHV\rangle\pm |VVH\rangle\right),\nonumber\\
|GHZ_3^\pm\rangle&=&\frac{1}{\sqrt{2}}\left(|HVH\rangle\pm |VHV\rangle\right),\nonumber\\
|GHZ_4^\pm\rangle&=&\frac{1}{\sqrt{2}}\left(|HVV\rangle\pm |VHH\rangle\right).
\end{eqnarray}

There are two cases for measuring photons with two key generation basis combinations.

Case 1: Using the first key generation basis combination $\{A_{1}B_{1}C_{1}\}$, the measurement results for the eight GHZ states can be written as
\begin{widetext}
\begin{eqnarray}\label{BGHZ4}
|GHZ_1^+\rangle&=&\frac{1}{2}(|+_x\rangle_a|+_x\rangle_b|+_x\rangle_c+|+_x\rangle_a|-_x\rangle_b|-_x\rangle_c
+|-_x\rangle_a|+_x\rangle_b|-_x\rangle_c+|-_x\rangle_a|-_x\rangle_b|+_x\rangle_c),\nonumber\\
|GHZ_2^+\rangle&=&\frac{1}{2}(|+_x\rangle_a|+_x\rangle_b|+_x\rangle_c-|+_x\rangle_a|-_x\rangle_b|-_x\rangle_c
-|-_x\rangle_a|+_x\rangle_b|-_x\rangle_c+|-_x\rangle_a|-_x\rangle_b|+_x\rangle_c),\nonumber\\
|GHZ_3^+\rangle&=&\frac{1}{2}(|+_x\rangle_a|+_x\rangle_b|+_x\rangle_c-|+_x\rangle_a|-_x\rangle_b|-_x\rangle_c
+|-_x\rangle_a|+_x\rangle_b|-_x\rangle_c-|-_x\rangle_a|-_x\rangle_b|+_x\rangle_c),\nonumber\\
|GHZ_4^+\rangle&=&\frac{1}{2}(|+_x\rangle_a|+_x\rangle_b|+_x\rangle_c+|+_x\rangle_a|-_x\rangle_b|-_x\rangle_c
-|-_x\rangle_a|+_x\rangle_b|-_x\rangle_c-|-_x\rangle_a|-_x\rangle_b|+_x\rangle_c),\nonumber\\
|GHZ_1^-\rangle&=&\frac{1}{2}(|+_x\rangle_a|-_x\rangle_b|+_x\rangle_c+|-_x\rangle_a|+_x\rangle_b|+_x\rangle_c
+|+_x\rangle_a|+_x\rangle_b|-_x\rangle_c+|-_x\rangle_a|-_x\rangle_b|-_x\rangle_c),\nonumber\\
|GHZ_2^-\rangle&=&\frac{1}{2}(|+_x\rangle_a|-_x\rangle_b|+_x\rangle_c+|-_x\rangle_a|+_x\rangle_b|+_x\rangle_c
-|+_x\rangle_a|+_x\rangle_b|-_x\rangle_c-|-_x\rangle_a|-_x\rangle_b|-_x\rangle_c),\nonumber
\end{eqnarray}
\begin{eqnarray}
|GHZ_3^-\rangle&=&\frac{1}{2}(-|+_x\rangle_a|-_x\rangle_b|+_x\rangle_c+|-_x\rangle_a|+_x\rangle_b|+_x\rangle_c
+|+_x\rangle_a|+_x\rangle_b|-_x\rangle_c-|-_x\rangle_a|-_x\rangle_b|-_x\rangle_c),\nonumber\\
|GHZ_4^-\rangle&=&\frac{1}{2}(|+_x\rangle_a|-_x\rangle_b|+_x\rangle_c-|-_x\rangle_a|+_x\rangle_b|+_x\rangle_c
+|+_x\rangle_a|+_x\rangle_b|-_x\rangle_c-|-_x\rangle_a|-_x\rangle_b|-_x\rangle_c).
\end{eqnarray}
\end{widetext}

According to the coding rules in Sec.~\ref{Section2.2}: $K_A=K_B\oplus K_C$, when the photon state degenerates to $|GHZ_1^-\rangle$, $|GHZ_2^-\rangle$, $|GHZ_3^-\rangle$, or $|GHZ_4^-\rangle$, three users use the key generation basis combination $\{A_1B_1C_1\}$ to produce the measurement results that would cause players (Bob and Charlie) to obtain the wrong key (\emph{i.e.} bit-flip errors have occurred). Therefore, for the first key generation basis combination $\{A_{1}B_{1}C_{1}\}$, the qubit error rate $Q_{111}$ in the white noise model is given by
\begin{eqnarray}\label{Q111}
Q_{111}=P(a_1b_1c_1\neq 1)=\frac{1}{8}(1-F)\times 4=\frac{1-F}{2}.
\end{eqnarray}

Case 2: Using the second key generation basis combination $\{A_{2}B_{1}C_{2}\}$, the measurement results for the eight GHZ states can be written as
\begin{widetext}
\begin{eqnarray}\label{BGHZ5}
|GHZ_1^+\rangle&=&\frac{1}{2}(|+_y\rangle_a|+_x\rangle_b|+_y\rangle_c+|+_y\rangle_a|-_x\rangle_b|-_y\rangle_c
+|-_y\rangle_a|+_x\rangle_b|-_y\rangle_c+|-_y\rangle_a|-_x\rangle_b|+_y\rangle_c),\nonumber\\
|GHZ_2^+\rangle&=&\frac{-i}{2}(|+_y\rangle_a|+_x\rangle_b|-_y\rangle_c-|+_y\rangle_a|-_x\rangle_b|+_y\rangle_c
-|-_y\rangle_a|+_x\rangle_b|+_y\rangle_c+|-_y\rangle_a|-_x\rangle_b|-_y\rangle_c),\nonumber\\
|GHZ_3^+\rangle&=&\frac{1}{2}(|+_y\rangle_a|+_x\rangle_b|+_y\rangle_c-|+_y\rangle_a|-_x\rangle_b|-_y\rangle_c
+|-_y\rangle_a|+_x\rangle_b|-_y\rangle_c-|-_y\rangle_a|-_x\rangle_b|+_y\rangle_c),\nonumber\\
|GHZ_4^+\rangle&=&\frac{-i}{2}(|+_y\rangle_a|+_x\rangle_b|-_y\rangle_c+|+_y\rangle_a|-_x\rangle_b|+_y\rangle_c
-|-_y\rangle_a|+_x\rangle_b|+_y\rangle_c-|-_y\rangle_a|-_x\rangle_b|+_y\rangle_c),\nonumber\\
|GHZ_1^-\rangle&=&\frac{1}{2}(|+_y\rangle_a|+_x\rangle_b|-_y\rangle_c+|+_y\rangle_a|-_x\rangle_b|+_y\rangle_c
+|-_y\rangle_a|+_x\rangle_b|+_y\rangle_c+|-_y\rangle_a|-_x\rangle_b|-_y\rangle_c),\nonumber\\
|GHZ_2^-\rangle&=&\frac{-i}{2}(-|+_y\rangle_a|+_x\rangle_b|+_y\rangle_c+|+_y\rangle_a|-_x\rangle_b|-_y\rangle_c
+|-_y\rangle_a|+_x\rangle_b|-_y\rangle_c-|-_y\rangle_a|-_x\rangle_b|+_y\rangle_c),\nonumber\\
|GHZ_3^-\rangle&=&\frac{1}{2}(|+_y\rangle_a|+_x\rangle_b|-_y\rangle_c-|+_y\rangle_a|-_x\rangle_b|+_y\rangle_c
+|-_y\rangle_a|+_x\rangle_b|+_y\rangle_c-|-_y\rangle_a|-_x\rangle_b|-_y\rangle_c),\nonumber\\
|GHZ_4^-\rangle&=&\frac{-i}{2}(|+_y\rangle_a|+_x\rangle_b|+_y\rangle_c+|+_y\rangle_a|-_x\rangle_b|-_y\rangle_c
-|-_y\rangle_a|+_x\rangle_b|-_y\rangle_c-|-_y\rangle_a|-_x\rangle_b|+_y\rangle_c).
\end{eqnarray}
\end{widetext}

According to the coding rules in Sec.~\ref{Section2.2}, when the photon state degenerates to $|GHZ_1^-\rangle$, $|GHZ_2^+\rangle$, $|GHZ_3^-\rangle$, or $|GHZ_4^+\rangle$, the measurement results will cause Bob to obtain wrong key. So, for the second key generation basis combination $\{A_{2}B_{1}C_{2}\}$, the QBER $Q_{212}$ in the white noise model is given by
\begin{eqnarray}\label{Q212}
Q_{212}=P(a_2b_1c_2\neq 1)=\frac{1}{8}(1-F)\times 4=\frac{1-F}{2}.
\end{eqnarray}

In conclusion, the measurement results of the two key generation bases have same QBER in the white noise model as
\begin{eqnarray}\label{Q1}
Q_1=Q_{111}=Q_{212}=\frac{1-F}{2}.
\end{eqnarray}

\end{document}